\documentclass[aps,prl,amsmath,amssymb,twocolumn]{revtex4-2}

\usepackage[T1]{fontenc} % Use modern font encodings
\usepackage{float}
\usepackage{dcolumn}
\usepackage{graphicx}
\usepackage{siunitx}
\usepackage{amsmath}
\usepackage{amssymb}
\usepackage{chemformula}
\usepackage{braket}
\usepackage{float}
\usepackage{graphicx}
\usepackage{siunitx}
\usepackage{amsfonts}
\usepackage{color}
\usepackage{soul}
\usepackage{braket}
\usepackage{verbatim}
\usepackage{tabu}

\begin{document}
	
\title{Temperature dependence of Coherent versus spontaneous Raman Scattering}
	
	\author{%
		Giovanni Batignani$^{1,2}$,
		Emanuele Mai$^{1,2}$,
		Miles Martinati$^{1}$,
		Mohanan M. Neethish$^{1}$,
		Shaul Mukamel$^{3}$,
		Tullio Scopigno$^{1,2,4}$,
	}
	
	\affiliation{$^{1}$ Dipartimento di Fisica,~Universit\'a~di~Roma~``La Sapienza'',  ~Roma, ~I-00185, ~Italy}
	\affiliation{$^{2}$ Center for Life Nano Science @Sapienza, Istituto Italiano di Tecnologia, Viale Regina Elena 291, I-00161, Roma,Italy}
	\affiliation{$^{3}$ Department of Chemistry, University of California, Irvine, 92623, California, USA}
	\affiliation{$^{4}$ Istituto Italiano di Tecnologia, Graphene Labs, Via Morego 30, I-16163 Genova, Italy}
%	\affiliation{$^{*}$ giovanni.batignani@uniroma1.it, $^{\dagger}$ tullio.scopigno@uniroma1.it}
	\email{giovanni.batignani@uniroma1.it and tullio.scopigno@uniroma1.it}
	
	\date{\today}
	
\begin{abstract}
Due to their sub picosecond temporal resolution, coherent Raman spectroscopies have been proposed as a viable extension of Spontaneous Raman (SR) thermometry, to determine dynamics of mode specific vibrational energy content during out of equilibrium molecular processes. Here we show that the presence of multiple laser fields stimulating the vibrational coherences introduces additional quantum pathways, resulting in destructive interference. This ultimately reduces the thermal sensitivity of single spectral lines, nullifying it for harmonic vibrations and temperature independent polarizability. We demonstrate how harnessing anharmonic signatures such as vibrational hot bands enables coherent Raman thermometry.
\end{abstract}

\maketitle

Ever since the Raman effect was first discovered, the classical description of the scattering process has emphasized the temperature dependence of the Raman cross section, being activated through the polarizability modulations generated by the structural fluctuations with respect to the equilibrium state~\cite{Raman_1928}. While this classical picture captures the essence of the physical process by predicting larger Raman signals at higher temperatures, it fails to provide a quantitative description of the measured responses.
Indeed, it does not account for the temperature-dependent Stokes and anti-Stokes line intensities, nor does it predict a non-vanishing (Stokes) response as the temperature approaches absolute 0 K. This discrepancy can be attributed to the low occupation numbers typically associated with thermal excitations~\cite{Cardona_book}, revealing the quantum nature of vibrational levels.
Under low light excitation regimes~\cite{Kasperczyk2015}, the ratio of Stokes ($I_{S}$) and anti-Stokes ($I_{AS}$) line intensities is linked to the temperature by the equation: $\frac{I_{S}}{I_{AS}} = \frac{\nu_S^3}{\nu_{AS}^3} e^{\frac{\hbar \omega_0}{k_B T}}$, with $\nu_S=\nu_R-\nu_0$ and $\nu_{AS}=\nu_R+\nu_0$ indicating the frequencies of Stokes and anti-Stokes emissions.
This serves as a powerful non-contact tool for extracting the equilibrium temperature of the sample under investigation, being routinely applied for the characterization of both gas, flames, liquid, solid state compounds~\cite{Sandell2020,Kumar2022}, as well as for studying relaxation processes~\cite{Mizutani1997, Ye2003,Keller2018} and thermal transport~\cite{Soini2010,Braun2022}. 	
	
Early observation of an asymmetry between the vibrational response in the red and in the blue side of the spectrum (corresponding to lower and higher frequency probe pulses relative to the pump) measured by stimulated Raman spectroscopy experiments (SRS)~\cite{Uma2011} has been interpreted in terms of Stokes/anti-Stokes temperature dependent ratio, in line with the SR scenario, advancing a straightforward application of SRS for  thermometry~\cite{Dang_2011}. Theoretical contributions~\cite{Mukamel_2013} clarified that, in striking contrast with SR, the same (thermal) populations concur to the generation of red and blue shifted SRS components, which hence cannot be compared to extract the temperature. 
	 
\begin{figure*}
		\centering
		\includegraphics[width=11 cm]{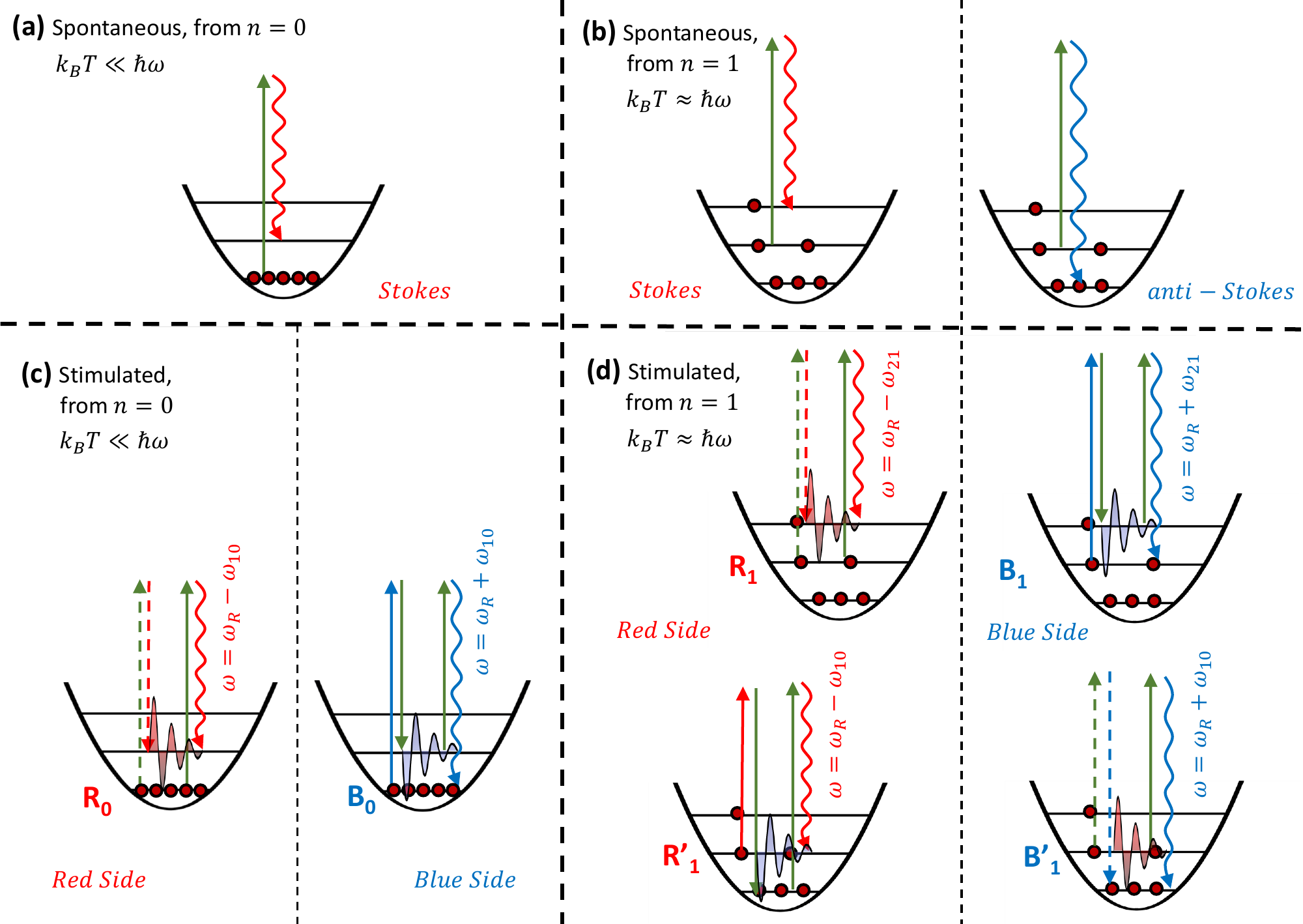}
		\caption{Energy levels involved in Stokes/anti-Stokes SR processes from the ground/first excited vibrational levels (a-b). 
		The corresponding pathways, describing the molecular density matrix evolution and the interaction with the electromagnetic fields, are shown in (c-d) for off-resonant SRS processes; dashed/solid lines indicate interactions with the bra/ket. For a molecule initially in the $n=0$ level, two pathways ($R_0$ and $B_0$) concur to the generation of the SRS response, generating a peak/loss in the red/blue side. When the molecule is initially in the $n=1$ state, additional pathways are enabled: $R_1$/$B_1$ are the natural extension of $R_0$/$B_0$ and probe the 
		$\ket{1}\rightarrow \ket{2}$ transition, while $R'_1$/$B'_1$ involve the $\ket{0}\rightarrow \ket{1}$ transition.
		Spontaneos Stokes pathways correspond to $R_i$ diagrams (leading to gains in the red side), whereas spontaneos anti-Stokes processes originate from $B'_i$ diagrams (blue side gains). In both cases the vacuum field replaces the probe beam.
		}\label{Fig1}
\end{figure*}

Here, we build on a diagrammatic framework to dissect the different processes that concur to the generation of nonlinear Raman responses~\cite{Lee2004,Dorfman2013}, studying how the thermal distribution affects the measured signals.
%Stimulated Raman spectroscopy is a powerful technique for studying the vibrational properties of molecular and solid-state systems~\cite{Mukamel_Potma_CRS,Prince2016}. By combining a narrowband Raman pump (RP) and broadband probe pulse (PP), it can be harnessed to map the Raman response over a large spectral range: 
SRS combines a narrowband Raman pump (RP) and broadband probe pulse (PP) to access the vibrational spectrum~\cite{Mukamel_Potma_CRS,Prince2016}:
the interaction of the sample with PP spectral components shifted by one vibrational quantum with respect to the RP  stimulates vibrational coherences, which in turn modulate the macroscopic polarizability, generating Raman bands on top of the probe spectrum~\cite{Mathies_review,Ferrante2018}. 
Coherent Raman spectroscopy (CRS) is also exploited to probe vibrational excitations directly in the time-domain~\cite{Ruhman1988,Kuramochi2021,Monacelli2017,Batignani2022}, by using a femtosecond RP to stimulate vibrational coherences, which are then recorded by scanning the time-dependent transmissivity via a broadband PP~\cite{cit::IVS::kukura,Gdor2017,Batignani_2021}. This scheme is conventionally referred to as impulsive stimulated Raman scattering (ISRS). Adding a femtosecond Actinic pump, which temporally precedes the RP-PP pair, turns SRS and ISRS into powerful pump-probe schemes~\cite{Yoshizawa1994,Yoshizawa1999, Takeuchi2008,Dietze2016}, which, by combining high spectral resolution and temporal precision~\cite{Biggs2011, Kuramochi2021}, represent ideal tools to
track structural changes in ultrafast photochemical and photophysical processes~\cite{Provencher2014,Batignani2015,Zhou2015,Batignani2016,Barclay2017,Quincy2018,Takaya2018,Hontani2018,Piontkowski2018,Fang2019,Fujisawa2016,Kim2020,Kim2021,BorregoVarillas2021}. 
\\
In order to have a closer comparison with spontaneous Raman spectroscopy, we focus on frequency-domain SRS. The diagrams describing the radiation-matter interactions responsible for the signal generation of the former, indeed, coincides with a subset of the latter's.  SRS experiments are sensitive to the third-order susceptibility, which is probed via the heterodyne detection of the induced nonlinear polarization $P^{(3)}$. Under the electronically off-resonant regime, i.e. when the pulses' wavelengths are detuned with respect to the sample absorption, the radiation-matter interaction Hamiltonian involved in the preparation and detection of the vibrational coherences is~\cite{Tanimura1993} $H_{I}^{(RP)} = -\alpha \cdot |E_{RP}|^2$, where $\alpha$ is the  molecular polarizability.
The Raman response can be isolated from the spectrally resolved probe via the differential Raman gain $RG=I_P(\omega)/I^0_P(\omega)-1$, where $I_P(\omega)$ ($I^0_P(\omega)$) indicates the PP spectrum recorded in presence (absence) of the RP. 
In Fig.~\ref{Fig1}c, the pathways responsible for the SRS signal are reported for both the red ($R_0$ diagram) and the blue ($B_0$ diagram) side of the spectrum, considering a system initially in the ground state ($n = 0$). 
Briefly, in the $R_0$ process an interaction between the bra side of the density matrix and the RP is followed by an interaction with the PP, which brings the system to the $\ket{0}\bra{1}$ vibrational coherence. Then an additional interaction with the RP occurs on the ket side and precedes the free induction decay that leaves the system in the $\ket{1}\bra{1}$  population.
In the $B_0$ process all the interactions occur on the ket side: the vibrational coherence is prepared by an interaction with a PP spectral component blue-shifted with respect to the pump, degenerate in frequency with the emission. Under the nonresonant regime, such processes give rise to the following polarizabilities~\cite{Batignani2016SciRep,Batignani2020}
	\begin{equation}\label{Eq:PR0}
		P_{R_0}(\omega) = - \frac{P_0}{\hbar}\left[ \frac{\partial \alpha(T)}{\partial Q}\right]^2 |\bra{1}\hat{Q}\ket{0}|^2\frac{|E_R|^2 E_P}{\omega_R-\omega_{10}-\omega-i\gamma_{10}}
	\end{equation}
	\begin{equation}\label{Eq:PB0}
		P_{B_0}(\omega) = \frac{P_0}{\hbar}\left[ \frac{\partial \alpha(T)}{\partial Q}\right]^2 |\bra{1}\hat{Q}\ket{0}|^2\frac{|E_R|^2 E_P}{\omega_R+\omega_{10}-\omega-i\gamma_{10}}
	\end{equation}
where $\gamma_{ij}$ indicates the dephasing rate of the $\ket{i}\bra{j}$ coherence, $\omega_{ij}=\omega_{i}-\omega_{j}$, $P_j$ is the initial population in the $j$ state, $\frac{\partial \alpha}{\partial Q}$ is the molecular polarizability derivative with respect to the considered normal coordinate $Q$, $E_{R/P}$ denotes the RP/PP field amplitude. 
Without loss of generality, we have considered an impulsive PP and a monochromatic RP.
These polarizabilities can be exploited to compute the heterodyne detected differential RG via the relation~\cite{cit::Mukamel}
	$RG \propto -\Im \left[\omega P(\omega)/E_P(\omega)\right]$, valid in the low excitation regime~\cite{batignani_OMX}:
	$$
	 RG_{R_0}(\omega) \propto P_0\left[ \frac{\partial \alpha(T)}{\partial Q}\right]^2 \frac{|\bra{1}\hat{Q}\ket{0}|^2}{\hbar}\frac{\omega \gamma_{10} |E_R|^2}{\left(\omega_R-\omega_{10}-\omega\right)^2+\gamma_{10}^2}
	$$
	$$
	 RG_{B_0}(\omega) \propto - P_0\left[ \frac{\partial \alpha(T)}{\partial Q}\right]^2 
	\frac{|\bra{1}\hat{Q}\ket{0}|^2}{\hbar}
	\frac{\omega \gamma_{10}|E_R|^2 }{(\omega_R+\omega_{10}-\omega)^2+\gamma_{10}^2}
	$$
$RG_{R_0}(\omega)$ corresponds to a positive gain in the red side (at $\omega=\omega_R-\omega_{10}$), while $ RG_{B_0}(\omega)$ generates a negative loss in the blue side (at $\omega=\omega_R+\omega_{10}$). The amplitude of both the signals is proportional to the polarizability derivative square modulus, the RP intensity $I_R=|E_R|^2$ and the $|\bra{1}\hat{Q}\ket{0}|^2$ wavefunctions overlap.
	
When the sample temperature increases, vibrational levels with an occupation number greater than 0 can be initially populated, enabling additional Raman transitions and potentially temperature dependent polarizability. An accurate calculation of the response function requires hence summing over all the contributions from thermally populated higher-lying levels.
For instance, SRS probes $n=1 \rightarrow n=2$ transitions via the following molecular susceptibility
	$$
	P_{R_1}(\omega) = - \frac{P_1}{\hbar}\left[ \frac{\partial \alpha(T)}{\partial Q}\right]^2 |\bra{2}\hat{Q}\ket{1}|^2\frac{|E_R|^2 E_P}{\omega_R-\omega_{21}-\omega-i\gamma_{21}}
	$$
	%\end{equation}
	which is formally analogous to Eq.~\ref{Eq:PR0} and gives rise to a positive peak at $\omega=\omega_R-\omega_{21}$. 
	Similarly, in the blue side $n=1 \rightarrow n=2$ transitions are probed by the following third-order polarizability:
	%\begin{equation}\label{Eq:PB1}
	$$
	P_{B_1}(\omega) = \frac{P_1}{\hbar}\left[ \frac{\partial \alpha(T)}{\partial Q}\right]^2 |\bra{2}\hat{Q}\ket{1}|^2\frac{|E_R|^2 E_P}{\omega_R+\omega_{21}-\omega-i\gamma_{21}}
	$$
	
Critically, permutations of the RP and PP fields are now possible giving also rise to the $R_1'$ and $B_1'$ pathways depicted in Fig.~\ref{Fig1}d, involving the $\ket{0}\bra{1}$ and $\ket{1}\bra{0}$ vibrational coherences, respectively.
%In the $R_1'$ process an interaction between the ket side of the density matrix and the PP is followed by an interaction with the RP, which leads the system to the $\ket{0}\bra{1}$ vibrational coherence. Then an additional interaction with the RP occurs on the ket side and precedes the free induction decay that leaves the system in the $\ket{1}\bra{1}$  population. 
%For the $B_1'$ process, the $\ket{1}\bra{0}$ vibrational coherence is stimulated by two interactions on the bra side with the RP-PP pair, being probed via an interaction on the ket with the RP before the free induction decay. 
The corresponding polarizabilities read as
	$$
	P_{R'_1}(\omega) =  \frac{P_1}{\hbar}\left[ \frac{\partial \alpha(T)}{\partial Q}\right]^2 |\bra{1}\hat{Q}\ket{0}|^2\frac{|E_R|^2 E_P}{\omega_R-\omega_{10}-\omega-i\gamma_{10}}
	$$
	$$
	P_{B'_1}(\omega)  = - \frac{P_1}{\hbar}\left[ \frac{\partial \alpha(T)}{\partial Q}\right]^2 |\bra{1}\hat{Q}\ket{0}|^2\frac{|E_R|^2 E_P}{\omega_R+\omega_{10}-\omega-i\gamma_{10}}
	$$
	
Interestingly, the molecular susceptibilities appearing in the $R_1'$-$B_1'$ responses are identical to those observed in the $R_0$-$B_0$ pathways. However, the $R_1'$-$B_1'$ intensities vary in amplitude (being directly proportional to the initial $P_1$ population) and, most importantly, exhibit opposite signs, leading to a destructive interference with the $R_0$-$B_0$ pathways. This can be rationalized as an effect analogous to the competition between stimulated emission and absorption processes in a two level system~\cite{cit::LoudonBook}.
\\	
The measured SRS response can be finally derived by summing over the initial vibrational levels, weighted by the corresponding thermal populations. The resulting red/blue side RGs read:	
	\begin{multline}\label{SRS_Red_Summed}
		 RG_{R}(\omega) \propto \left[ \frac{\partial \alpha(T)}{\partial Q}\right]^2 \omega I_R 
		\sum_{j=0}^\infty (P_j-P_{j+1}) \\
		\frac{ \gamma_j |\bra{j+1}\hat{Q}\ket{j}|^2}{\left(\omega_R-\omega_j-\omega\right)^2+\gamma_j^2}
	\end{multline}
	\begin{multline}\label{SRS_Blue_Summed}
		 RG_{B}(\omega) \propto - \left[ \frac{\partial \alpha(T)}{\partial Q}\right]^2 \omega I_R 
		\sum_{j=0}^\infty (P_j-P_{j+1}) \\ \frac{ \gamma_j |\bra{j+1}\hat{Q}\ket{j}|^2}{\left(\omega_R+\omega_j-\omega\right)^2+\gamma_j^2}
	\end{multline}
where $P_j=\frac{e^{-E_j/{k_BT}}}{Z}$ and $Z=\sum_0^{\infty}e^{-E_j/{k_BT}}$ is the partition function. For improved readability, we have indicated $\omega_{j+1,j}$ and $\gamma_{j+1,j}$ as $\omega_j$ and $\gamma_j$. 
Equations~\ref{SRS_Red_Summed}-\ref{SRS_Blue_Summed} show the same dependence on the vibrational population, indicating that the ratio between the red/blue side SRS responses is independent of temperature, rationalizing the results of Ref.~\cite{Mukamel_2013} in terms of sum over the vibrational levels.
\\
It is worth to stress that this result is valid also for large excitation regimes: high RP fluences can indeed result in larger RGs, once the exponential gain regime is initiated \cite{batignani_OMX}. However, they do not alter the SRS sensitivity to the molecular properties, as the measured signal still depends on the same third-order polarization as $ RG_{R/B}(\omega)=e^{-\frac{2\pi \omega \Im \left[ P^{(3)}_{tot}(\omega) / E_P(\omega) \right]  L}{c n }}-1$, where $n$ and $L$ indicate the sample refractive index and length, respectively. 
	
The spontaneous Raman case can be seen as a SRS process in which the probe pulse is provided by the vacuum field.  Since this latter cannot be annihilated, only those pathways which contain two matter de-excitations into the vacuum field contributes to SR, corresponding to the diagrams where the last two interactions are complex conjugate of the first two ($R_i$ and $B'_i$).
\\
This results in the well known Stokes (red side, with final level higher than the initial one) and anti-Stokes (blue side, with final state lower than the initial one) processes, naturally weighted by different thermal populations. 
The summation over all the possible directions of the scattered field leads to a cubic dependence of the spontaneous Raman cross section on the emitted frequency $\omega$, in contrast with the linear dependence ruling stimulated processes~\cite{cit::LoudonBook}; the corresponding Stokes/anti-Stokes signals read:
	\begin{equation}\label{Eq:Spont_with_sum}
		\begin{cases}
			\begin{split}
		R_{S}(\omega) \propto \left[\frac{\partial \alpha(T)}{\partial Q}\right]^2 \omega^3 I_R
		\sum_{j=0}^\infty P_j\frac{ \gamma_j |\bra{j+1}\hat{Q}\ket{j}|^2}{\left(\omega_R-\omega_j-\omega\right)^2+\gamma_j^2}
		\\
		R_{AS}(\omega) \propto \left[\frac{\partial \alpha(T)}{\partial Q}\right]^2 \omega^3  I_R
		\sum_{j=0}^\infty P_{j+1}\frac{\gamma_j |\bra{j}\hat{Q}\ket{j+1}|^2}{\left(\omega_R+\omega_j-\omega\right)^2+\gamma_j^2}
	\end{split}
\end{cases}
	\end{equation}
	
Summing up, for a given normal mode, both SRS and SR emit on the red and blue side depending on the relative wavelengths of the last two interactions, which scatter off the vibrational coherence. Critically, in SR the two cases can only originate from Stokes and anti Stokes processes, respectively: therefore the measured response depends directly on the temperature via the thermal populations. In the stimulated regime this is no longer true: two Stokes contributions, originating from the coherently driven vibrational coherences,  can be scattered off by different processes emitting on the red or blue side (R and B families in Fig. \ref{Fig1}).    

All the aforementioned  considerations can also be applied to coherent Stokes/anti-Stokes Raman scattering and ISRS,  with the remarkable distinction that the vibrational coherence preparation and the pulses  are different~\cite{Monacelli2017,Polli2018,Virga2019}. The derivation of the corresponding polarizations is reported in SI.

To assess the effect of thermal excitations on spontaneous versus coherent Raman spectra, we start considering a simple harmonic molecular system, with frequency $\omega_v$ and a $j$-independent dephasing rate ($\gamma_j=\gamma_v$), evaluating the $\sum_{j=0}^\infty P_j$ summation. The thermal occupation numbers of the $j$-levels are
	$
	P_j=\left(e^{-j \hbar \omega_v/(k_BT)}\right)\cdot \left(1-e^{-\hbar \omega_v/(k_BT)}\right)
	$, while 
\begin{equation}\label{eq:deltaP}
\Delta P_j=P_j-P_{j+1} = e^{-j \hbar \omega_v/(k_BT)}\cdot \left(1-e^{-\hbar \omega_v/(k_BT)}\right)^2
\end{equation}
Finally, the transition amplitude $\bra{j+1}\hat{Q}\ket{j}$ reads
	\begin{equation}
		\bra{j+1}\hat{Q}\ket{j}=\bra{j+1}\sqrt{\frac{\hbar}{2m \omega_v}}\left( \hat{a}+\hat{a}^{\dagger}\right)\ket{j}=
		\sqrt{\frac{(j+1) \hbar}{2m \omega_v}}
	\end{equation}
	which increases as the initial occupation number $j$ is increased, critically determining the temperature dependence of spontaneous Raman responses.  
	Plugging these relations into Eq.~\ref{Eq:Spont_with_sum}, the spontaneous Stokes response can be expressed as
	\begin{multline*}
		R_{S}(\omega) \propto \left[\frac{\partial \alpha(T)}{\partial Q}\right]^2 
		\frac{ \gamma_j \omega^3 I_R}{\left(\omega_R-\omega_v-\omega\right)^2+\gamma_v^2}\cdot
		\\
		\cdot \frac{\hbar}{2m \omega_v}
		\sum_{j=0}^\infty \left(j+1\right) \cdot e^{-j \hbar \omega_v/(k_BT)}\cdot \left(1-e^{-\hbar \omega_v/(k_BT)}\right)  
	\end{multline*}
Carrying out the summation over the initial states $j$ and incorporating the constant $\frac{\hbar}{2m \omega_v}$ term into the proportionality relation, $R_{S}(\omega)$ reads
	\begin{equation}\label{Eq:Stokes}
		R_{S}(\omega) \propto \left[\frac{\partial \alpha(T)}{\partial Q}\right]^2 
		\frac{ \gamma_j \omega^3 I_R}{\left(\omega_R-\omega_v-\omega\right)^2+\gamma_v^2}\cdot
		\Big( \langle n \rangle +1\Big)
	\end{equation}
where $\langle n \rangle = \frac{1}{e^{\hbar \omega_v/(k_BT)}-1}$ is the mean occupation number. Similarly, $R_{AS}(\omega)$ reads
	\begin{equation}\label{Eq:anti-Stokes}
		R_{AS}(\omega) \propto \left[\frac{\partial \alpha(T)}{\partial Q}\right]^2 
		\frac{ \gamma_j \omega^3 I_R}{\left(\omega_R+\omega_v-\omega\right)^2+\gamma_v^2}\cdot
		\langle n \rangle
	\end{equation}
	
On the other hand, the temperature dependence of the red/blue side SRS responses is ruled by the following summation
	\begin{multline*}
		\sum_{j=0}^\infty (P_j-P_{j+1}) \cdot (j+1) = \\
		\sum_{j=0}^\infty
		(j+1) \cdot e^{-j \hbar \omega_v/(k_BT)}\cdot \left(1-e^{-\hbar \omega_v/(k_BT)}\right)^2= 
		1
	\end{multline*}
Therefore, the Raman gains on the red/blue side can be expressed as
	\begin{equation}\label{Eq:Red_T-independent}
		 RG_{R}(\omega) \propto \left[ \frac{\partial \alpha(T)}{\partial Q}\right]^2 \cdot
		\frac{\gamma_v \omega I_R }{\left(\omega_R-\omega_v-\omega\right)^2+\gamma_v^2}
	\end{equation}
	\begin{equation}\label{Eq:Blue_T-independent}
		 RG_{B}(\omega) \propto - \left[ \frac{\partial \alpha(T)}{\partial Q}\right]^2 \cdot
		\frac{ \gamma_v \omega I_R}{\left(\omega_R+\omega_v-\omega\right)^2+\gamma_v^2}
	\end{equation}
notably, the only temperature dependence is brought about by the polarizability gradient.
%A comparison of the SR/SRS temperature-dependence computed using Eqs.~\ref{Eq:Stokes}~-~\ref{Eq:Blue_T-independent} is reported in Fig.~\ref{Fig2}a-b, for an ideal harmonic system with 3 normal modes.
%%The SRS signal temperature-dependence computed using Eqs.~\ref{Eq:Red_T-independent}~-~\ref{Eq:Blue_T-independent} is reported in Fig.~\ref{Fig2}a-b for an ideal harmonic system with 3 normal modes and is compared with the SR response. 

\begin{figure}
	\includegraphics[width=9 cm]{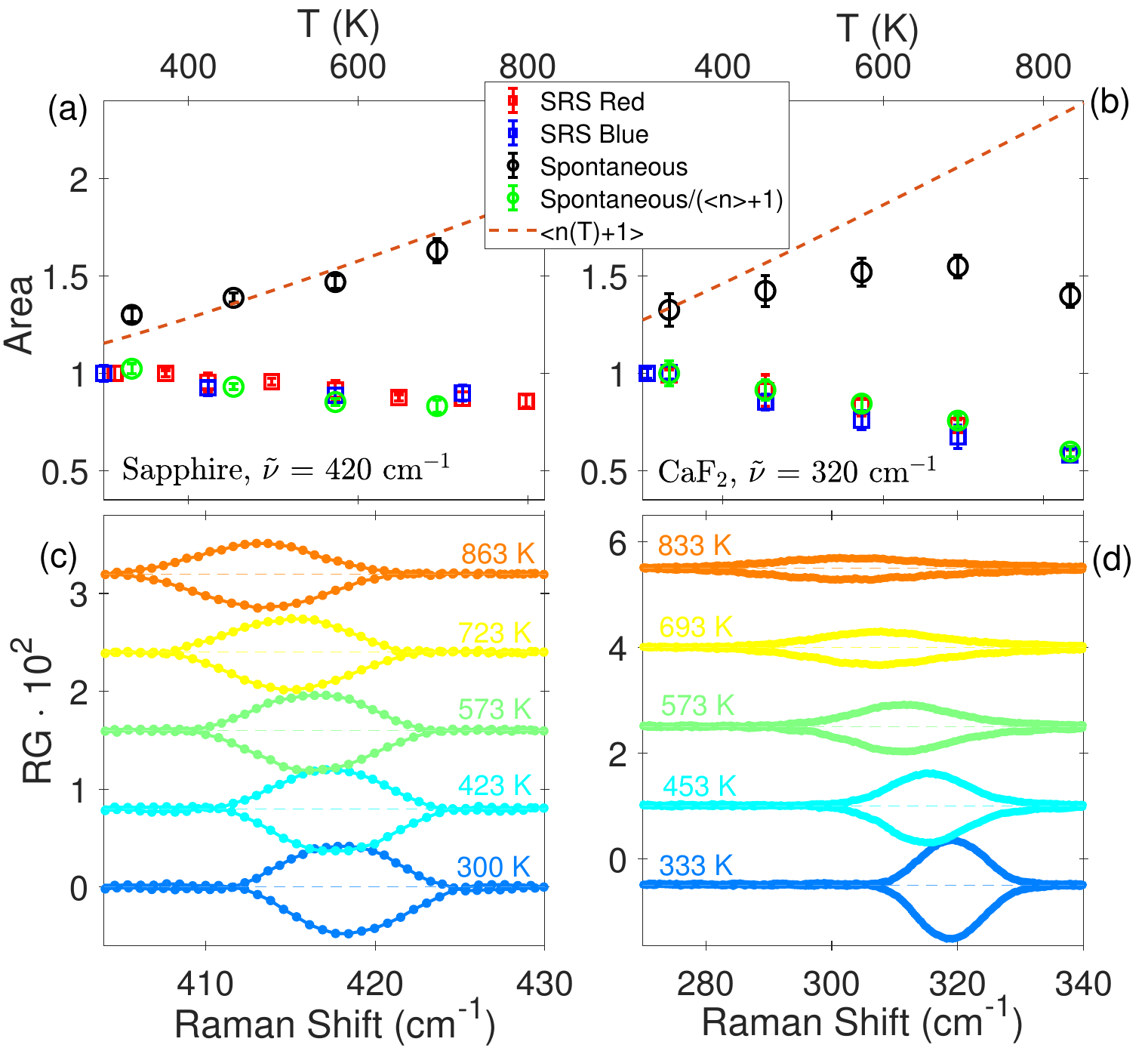}
	\caption{Temperature dependence of spontaneous and stimulated Raman responses, in the red and in the blue sides. Experimental responses have been measured for two crystals, namely Sapphire (vibrational band at $\sim$ 420 cm$^{-1}$) and \ch{CaF_2} (at  $\sim$ 320 cm$^{-1}$), and are reported evaluating the peak areas in panels a-b, respectively. Panels c-d show the corresponding SRS spectra, measured in the red (gains) and in the blue (losses) sides.
	}\label{FigR}
\end{figure}

To benchmark the presented theoretical results, SRS spectra have been measured for two crystals, namely Sapphire and \ch{CaF_2}, as function of temperature. The integrated area of the monitored bands (centred at $\sim$ 420 cm$^{-1}$ and $\sim$ 320 cm$^{-1}$, respectively) are reported in Fig.~\ref{FigR}. Notably, the areas of Sapphire and \ch{CaF_2} bands  experiences a reduction of approximately 10\% and 40\% respectively, as the temperature increases from 300K to 800K. To verify our theoretical prediction, we also measured the corresponding SR response over the same temperature range, which shows a significant deviation from the mean occupation number ($\langle n \rangle+1$). Critically, by factoring out such dependence we find an excellent agreement with the SRS data, validating our results. This also indicates that in the harmonic limit the observed decreasing trend is a measure of the decrease of the molecular polarizability derivative ($\frac{\partial \alpha(T)}{\partial Q}$).

%Indeed, the SR data normalized by the ($\langle n \rangle+1$) factor are in good agreement with the SRS intensities, validating such hypothesis. 
%These results suggest that SRS offers a convenient means to directly measure the $\frac{\partial \alpha(T)}{\partial Q}$ temperature dependence.

\begin{figure}
	\includegraphics[width=9 cm]{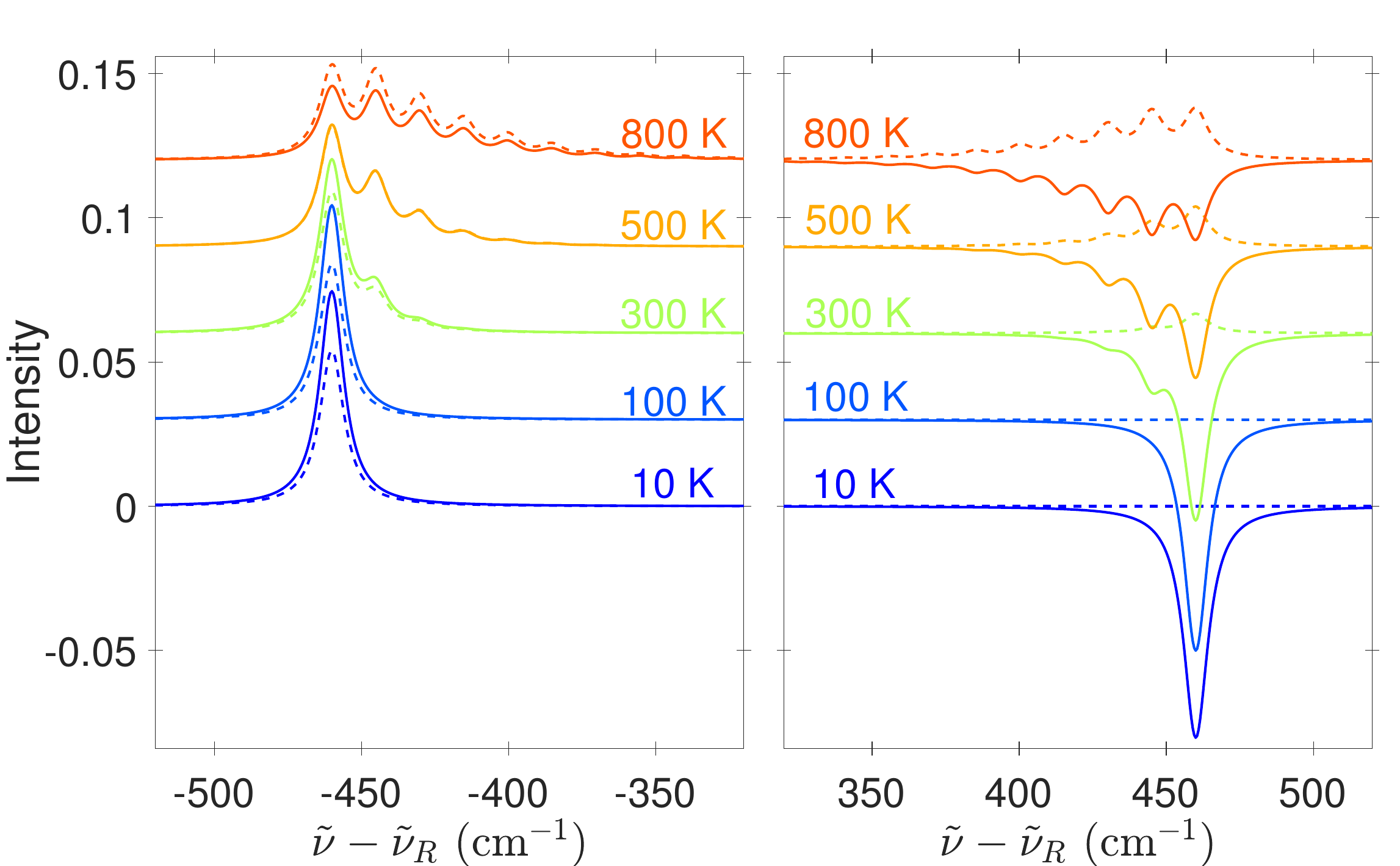}
	\caption{Comparison of stimulated vs spontaneous (solid/dashed lines) Raman responses as a function of temperature for a model system, with a single mode molecular anharmonicity. A Morse potential, with a fundamental band at 460 cm$^{-1}$, constant dephasing rate $\gamma_j=7$ cm$^{-1}$ and subsequent transitions shifted by 15 cm$^{-1}$, is considered. A monochromatic RP ($\lambda_{RP} = 800$ nm) and a spectrally flat PP are assumed.
	}\label{Fig3}
\end{figure}

The question now stands on whether coherent vibrational Raman can be used for thermometry. While it is well understood~\cite{Mukamel_2013} that the red/blue side ratio is thermally inactive, our result indicates that a single spectral component features a temperature dependence brought about by the sample polarizability, even in the harmonic case. Exploiting such a dependence would require either a calibration or an a priori knowledge of $\left[ \frac{\partial \alpha(T)}{\partial Q}\right]^2$. A convenient alternative is granted by anharmonic effects. A simple illustrative case is offered by single mode molecular anharmonicity, which is well captured by the Morse potential~\cite{Dahl1988,Ferrante2016,Ferrante2020} model. As shown in Fig.~\ref{Fig3}, upon increasing temperature, the molecular redistribution over the energy levels affects both the polarizability and the vibrational spectrum with the appearance of the hot band progression. The former can be factored out by comparing, for example, the relative amplitudes of two adjacent components, which, from eq.~\ref{eq:deltaP}, is $\Delta P_{j+1}/\Delta P_{j}=e^{-\hbar\omega_v/(k_BT)}$.

In conclusion, we have studied the temperature dependence of Raman processes by combining a quantum description of matter with a perturbative expansion of the molecular density matrix. This approach allowed us to dissect how different transitions within the vibrational manifold participate to the scattering mechanism, providing a direct comparison between spontaneous and stimulated processes. 
Generally, molecular potentials exhibit Raman transition amplitudes that increases with the occupation number  ($\bra{j} \hat{Q} \ket{j+1} \propto \sqrt{j+1}$, for harmonic potentials): this makes spontaneous Raman enhanced by thermal fluctuations. The same molecular transitions, with the same transition amplitudes, are involved in coherent Raman spectroscopy and hence, a dependence similar to SR may be erroneously expected also for CRS.
However, due to the different permutations of the external electromagnetic fields participating to the radiation-matter interactions, additional pathways concur to the generation of CRS signals.
In particular, stimulated processes originating from vibrationally excited energy levels can engage in destructive interference with transitions from the ground state or lower-lying energy levels, ultimately modifying the temperature dependence.
In particular, the total cross section of CRS processes are less sensitive to thermal effects with respect to SR. In the specific case of a simple harmonic molecular potential, the off-resonant CRS response merely reflects the temperature dependence of the polarizability gradient, in sharp contrast with SR. 
For these reasons, stimulated Raman-based vibrational thermometry requires more refined methodologies, relying either on an  a priori knowledge of the molecular polarizability or on the direct observation of anharmonic signatures.

\begin{acknowledgments}
This project has received funding from  the  Project ECS 0000024 Rome Technopole, - CUP  B83C22002820006,  NRP Mission 4 Component 2 Investment 1.5conclusio,  Funded by the European Union - NextGenerationEU (T.S.).
		G.B. and T.S. acknowledge the `Progetti di Ricerca Medi 2021', the `Progetti di Ricerca Medi 2022' grants by Sapienza~Universit\'a~di~Roma. E.M. and M. M. acknowledge the `Progetti di Avvio alla Ricerca 2022' grants by Sapienza~Universit\'a~di~Roma. S.M. acknowledges the support of NSF   through grant CHE-2246379.
		G.B. and T.S. acknowledge the support of EU HORIZON funding programme through grant GA-101120832.
\end{acknowledgments}	

%\bibliography{biblio5}

\begin{thebibliography}{56}%
	\makeatletter
	\providecommand \@ifxundefined [1]{%
		\@ifx{#1\undefined}
	}%
	\providecommand \@ifnum [1]{%
		\ifnum #1\expandafter \@firstoftwo
		\else \expandafter \@secondoftwo
		\fi
	}%
	\providecommand \@ifx [1]{%
		\ifx #1\expandafter \@firstoftwo
		\else \expandafter \@secondoftwo
		\fi
	}%
	\providecommand \natexlab [1]{#1}%
	\providecommand \enquote  [1]{``#1''}%
	\providecommand \bibnamefont  [1]{#1}%
	\providecommand \bibfnamefont [1]{#1}%
	\providecommand \citenamefont [1]{#1}%
	\providecommand \href@noop [0]{\@secondoftwo}%
	\providecommand \href [0]{\begingroup \@sanitize@url \@href}%
	\providecommand \@href[1]{\@@startlink{#1}\@@href}%
	\providecommand \@@href[1]{\endgroup#1\@@endlink}%
	\providecommand \@sanitize@url [0]{\catcode `\\12\catcode `\$12\catcode
		`\&12\catcode `\#12\catcode `\^12\catcode `\_12\catcode `\%12\relax}%
	\providecommand \@@startlink[1]{}%
	\providecommand \@@endlink[0]{}%
	\providecommand \url  [0]{\begingroup\@sanitize@url \@url }%
	\providecommand \@url [1]{\endgroup\@href {#1}{\urlprefix }}%
	\providecommand \urlprefix  [0]{URL }%
	\providecommand \Eprint [0]{\href }%
	\providecommand \doibase [0]{https://doi.org/}%
	\providecommand \selectlanguage [0]{\@gobble}%
	\providecommand \bibinfo  [0]{\@secondoftwo}%
	\providecommand \bibfield  [0]{\@secondoftwo}%
	\providecommand \translation [1]{[#1]}%
	\providecommand \BibitemOpen [0]{}%
	\providecommand \bibitemStop [0]{}%
	\providecommand \bibitemNoStop [0]{.\EOS\space}%
	\providecommand \EOS [0]{\spacefactor3000\relax}%
	\providecommand \BibitemShut  [1]{\csname bibitem#1\endcsname}%
	\let\auto@bib@innerbib\@empty
	%</preamble>
	\bibitem [{\citenamefont {Raman}\ and\ \citenamefont
		{Krishnan}(1928)}]{Raman_1928}%
	\BibitemOpen
	\bibfield  {author} {\bibinfo {author} {\bibfnamefont {C.}~\bibnamefont
			{Raman}}\ and\ \bibinfo {author} {\bibfnamefont {K.~S.}\ \bibnamefont
			{Krishnan}},\ }\bibfield  {title} {\bibinfo {title} {A new type of secondary
			radiation},\ }\href {https://doi.org/10.1038/121501c0} {\bibfield  {journal}
		{\bibinfo  {journal} {Nature}\ }\textbf {\bibinfo {volume} {121}},\ \bibinfo
		{pages} {501} (\bibinfo {year} {1928})}\BibitemShut {NoStop}%
	\bibitem [{\citenamefont {Cardona}\ and\ \citenamefont
		{Güntherodt}(1982)}]{Cardona_book}%
	\BibitemOpen
	\bibfield  {author} {\bibinfo {author} {\bibfnamefont {M.}~\bibnamefont
			{Cardona}}\ and\ \bibinfo {author} {\bibfnamefont {G.}~\bibnamefont
			{Güntherodt}},\ }\href@noop {} {\emph {\bibinfo {title} {Light Scattering in
				Solids {II}}}}\ (\bibinfo  {publisher} {Springer Berlin Heidelberg},\
	\bibinfo {address} {Berlin},\ \bibinfo {year} {1982})\BibitemShut {NoStop}%
	\bibitem [{\citenamefont {Kasperczyk}\ \emph {et~al.}(2015)\citenamefont
		{Kasperczyk}, \citenamefont {Jorio}, \citenamefont {Neu}, \citenamefont
		{Maletinsky},\ and\ \citenamefont {Novotny}}]{Kasperczyk2015}%
	\BibitemOpen
	\bibfield  {author} {\bibinfo {author} {\bibfnamefont {M.}~\bibnamefont
			{Kasperczyk}}, \bibinfo {author} {\bibfnamefont {A.}~\bibnamefont {Jorio}},
		\bibinfo {author} {\bibfnamefont {E.}~\bibnamefont {Neu}}, \bibinfo {author}
		{\bibfnamefont {P.}~\bibnamefont {Maletinsky}},\ and\ \bibinfo {author}
		{\bibfnamefont {L.}~\bibnamefont {Novotny}},\ }\bibfield  {title} {\bibinfo
		{title} {Stokes–anti-stokes correlations in diamond},\ }\href
	{https://doi.org/10.1364/OL.40.002393} {\bibfield  {journal} {\bibinfo
			{journal} {Opt. Lett.}\ }\textbf {\bibinfo {volume} {40}},\ \bibinfo {pages}
		{2393} (\bibinfo {year} {2015})}\BibitemShut {NoStop}%
	\bibitem [{\citenamefont {Sandell}\ \emph {et~al.}(2020)\citenamefont
		{Sandell}, \citenamefont {Chávez-Ángel}, \citenamefont {El~Sachat},
		\citenamefont {He}, \citenamefont {Sotomayor~Torres},\ and\ \citenamefont
		{Maire}}]{Sandell2020}%
	\BibitemOpen
	\bibfield  {author} {\bibinfo {author} {\bibfnamefont {S.}~\bibnamefont
			{Sandell}}, \bibinfo {author} {\bibfnamefont {E.}~\bibnamefont
			{Chávez-Ángel}}, \bibinfo {author} {\bibfnamefont {A.}~\bibnamefont
			{El~Sachat}}, \bibinfo {author} {\bibfnamefont {J.}~\bibnamefont {He}},
		\bibinfo {author} {\bibfnamefont {C.~M.}\ \bibnamefont {Sotomayor~Torres}},\
		and\ \bibinfo {author} {\bibfnamefont {J.}~\bibnamefont {Maire}},\ }\bibfield
	{title} {\bibinfo {title} {{Thermoreflectance techniques and Raman
				thermometry for thermal property characterization of nanostructures}},\
	}\href {https://doi.org/10.1063/5.0020239} {\bibfield  {journal} {\bibinfo
			{journal} {Journal of Applied Physics}\ }\textbf {\bibinfo {volume} {128}},\
		\bibinfo {pages} {131101} (\bibinfo {year} {2020})}\BibitemShut {NoStop}%
	\bibitem [{\citenamefont {Kumar}\ \emph {et~al.}(2022)\citenamefont {Kumar},
		\citenamefont {Stefanczyk}, \citenamefont {Chorazy}, \citenamefont
		{Nakabayashi},\ and\ \citenamefont {Ohkoshi}}]{Kumar2022}%
	\BibitemOpen
	\bibfield  {author} {\bibinfo {author} {\bibfnamefont {K.}~\bibnamefont
			{Kumar}}, \bibinfo {author} {\bibfnamefont {O.}~\bibnamefont {Stefanczyk}},
		\bibinfo {author} {\bibfnamefont {S.}~\bibnamefont {Chorazy}}, \bibinfo
		{author} {\bibfnamefont {K.}~\bibnamefont {Nakabayashi}},\ and\ \bibinfo
		{author} {\bibfnamefont {S.-i.}\ \bibnamefont {Ohkoshi}},\ }\bibfield
	{title} {\bibinfo {title} {Ratiometric raman and luminescent thermometers
			constructed from dysprosium thiocyanidometallate molecular magnets},\ }\href
	{https://doi.org/https://doi.org/10.1002/adom.202201675} {\bibfield
		{journal} {\bibinfo  {journal} {Advanced Optical Materials}\ }\textbf
		{\bibinfo {volume} {10}},\ \bibinfo {pages} {2201675} (\bibinfo {year}
		{2022})}\BibitemShut {NoStop}%
	\bibitem [{\citenamefont {Mizutani}(1997)}]{Mizutani1997}%
	\BibitemOpen
	\bibfield  {author} {\bibinfo {author} {\bibfnamefont {Y.}~\bibnamefont
			{Mizutani}},\ }\bibfield  {title} {\bibinfo {title} {Direct observation of
			cooling of heme upon photodissociation of carbonmonoxy myoglobin},\ }\href
	{https://doi.org/10.1126/science.278.5337.443} {\bibfield  {journal}
		{\bibinfo  {journal} {Science}\ }\textbf {\bibinfo {volume} {278}},\ \bibinfo
		{pages} {443} (\bibinfo {year} {1997})}\BibitemShut {NoStop}%
	\bibitem [{\citenamefont {Ye}\ \emph {et~al.}(2003)\citenamefont {Ye},
		\citenamefont {Demidov}, \citenamefont {Rosca}, \citenamefont {Wang},
		\citenamefont {Kumar}, \citenamefont {Ionascu}, \citenamefont {Zhu},
		\citenamefont {Barrick}, \citenamefont {Wharton},\ and\ \citenamefont
		{Champion}}]{Ye2003}%
	\BibitemOpen
	\bibfield  {author} {\bibinfo {author} {\bibfnamefont {X.}~\bibnamefont
			{Ye}}, \bibinfo {author} {\bibfnamefont {A.}~\bibnamefont {Demidov}},
		\bibinfo {author} {\bibfnamefont {F.}~\bibnamefont {Rosca}}, \bibinfo
		{author} {\bibfnamefont {W.}~\bibnamefont {Wang}}, \bibinfo {author}
		{\bibfnamefont {A.}~\bibnamefont {Kumar}}, \bibinfo {author} {\bibfnamefont
			{D.}~\bibnamefont {Ionascu}}, \bibinfo {author} {\bibfnamefont
			{L.}~\bibnamefont {Zhu}}, \bibinfo {author} {\bibfnamefont {D.}~\bibnamefont
			{Barrick}}, \bibinfo {author} {\bibfnamefont {D.}~\bibnamefont {Wharton}},\
		and\ \bibinfo {author} {\bibfnamefont {P.~M.}\ \bibnamefont {Champion}},\
	}\bibfield  {title} {\bibinfo {title} {Investigations of heme protein
			absorption line shapes, vibrational relaxation, and resonance raman
			scattering on ultrafast time scales},\ }\href
	{https://doi.org/10.1021/jp0276799} {\bibfield  {journal} {\bibinfo
			{journal} {The Journal of Physical Chemistry A}\ }\textbf {\bibinfo {volume}
			{107}},\ \bibinfo {pages} {8156} (\bibinfo {year} {2003})}\BibitemShut
	{NoStop}%
	\bibitem [{\citenamefont {Keller}\ and\ \citenamefont
		{Frontiera}(2018)}]{Keller2018}%
	\BibitemOpen
	\bibfield  {author} {\bibinfo {author} {\bibfnamefont {E.~L.}\ \bibnamefont
			{Keller}}\ and\ \bibinfo {author} {\bibfnamefont {R.~R.}\ \bibnamefont
			{Frontiera}},\ }\bibfield  {title} {\bibinfo {title} {Ultrafast nanoscale
			raman thermometry proves heating is not a primary mechanism for
			plasmon-driven photocatalysis},\ }\href
	{https://doi.org/10.1021/acsnano.8b01809} {\bibfield  {journal} {\bibinfo
			{journal} {ACS Nano}\ }\textbf {\bibinfo {volume} {12}},\ \bibinfo {pages}
		{5848} (\bibinfo {year} {2018})}\BibitemShut {NoStop}%
	\bibitem [{\citenamefont {Soini}\ \emph {et~al.}(2010)\citenamefont {Soini},
		\citenamefont {Zardo}, \citenamefont {Uccelli}, \citenamefont {Funk},
		\citenamefont {Koblmüller}, \citenamefont {Fontcuberta~i Morral},\ and\
		\citenamefont {Abstreiter}}]{Soini2010}%
	\BibitemOpen
	\bibfield  {author} {\bibinfo {author} {\bibfnamefont {M.}~\bibnamefont
			{Soini}}, \bibinfo {author} {\bibfnamefont {I.}~\bibnamefont {Zardo}},
		\bibinfo {author} {\bibfnamefont {E.}~\bibnamefont {Uccelli}}, \bibinfo
		{author} {\bibfnamefont {S.}~\bibnamefont {Funk}}, \bibinfo {author}
		{\bibfnamefont {G.}~\bibnamefont {Koblmüller}}, \bibinfo {author}
		{\bibfnamefont {A.}~\bibnamefont {Fontcuberta~i Morral}},\ and\ \bibinfo
		{author} {\bibfnamefont {G.}~\bibnamefont {Abstreiter}},\ }\bibfield  {title}
	{\bibinfo {title} {{Thermal conductivity of GaAs nanowires studied by
				micro-Raman spectroscopy combined with laser heating}},\ }\href
	{https://doi.org/10.1063/1.3532848} {\bibfield  {journal} {\bibinfo
			{journal} {Applied Physics Letters}\ }\textbf {\bibinfo {volume} {97}},\
		\bibinfo {pages} {263107} (\bibinfo {year} {2010})}\BibitemShut {NoStop}%
	\bibitem [{\citenamefont {Braun}\ \emph {et~al.}(2022)\citenamefont {Braun},
		\citenamefont {Furrer}, \citenamefont {Butti}, \citenamefont {Thodkar},
		\citenamefont {Shorubalko}, \citenamefont {Zardo}, \citenamefont {Calame},\
		and\ \citenamefont {Perrin}}]{Braun2022}%
	\BibitemOpen
	\bibfield  {author} {\bibinfo {author} {\bibfnamefont {O.}~\bibnamefont
			{Braun}}, \bibinfo {author} {\bibfnamefont {R.}~\bibnamefont {Furrer}},
		\bibinfo {author} {\bibfnamefont {P.}~\bibnamefont {Butti}}, \bibinfo
		{author} {\bibfnamefont {K.}~\bibnamefont {Thodkar}}, \bibinfo {author}
		{\bibfnamefont {I.}~\bibnamefont {Shorubalko}}, \bibinfo {author}
		{\bibfnamefont {I.}~\bibnamefont {Zardo}}, \bibinfo {author} {\bibfnamefont
			{M.}~\bibnamefont {Calame}},\ and\ \bibinfo {author} {\bibfnamefont {M.~L.}\
			\bibnamefont {Perrin}},\ }\bibfield  {title} {\bibinfo {title} {Spatially
			mapping thermal transport in graphene by an opto-thermal method},\ }\bibfield
	{journal} {\bibinfo  {journal} {npj 2D Materials and Applications}\ }\textbf
	{\bibinfo {volume} {6}},\ \href {https://doi.org/10.1038/s41699-021-00277-2}
	{10.1038/s41699-021-00277-2} (\bibinfo {year} {2022})\BibitemShut {NoStop}%
	\bibitem [{\citenamefont {Mallick}\ \emph {et~al.}(2011)\citenamefont
		{Mallick}, \citenamefont {Lakhsmanna},\ and\ \citenamefont
		{Umapathy}}]{Uma2011}%
	\BibitemOpen
	\bibfield  {author} {\bibinfo {author} {\bibfnamefont {B.}~\bibnamefont
			{Mallick}}, \bibinfo {author} {\bibfnamefont {A.}~\bibnamefont
			{Lakhsmanna}},\ and\ \bibinfo {author} {\bibfnamefont {S.}~\bibnamefont
			{Umapathy}},\ }\bibfield  {title} {\bibinfo {title} {Ultrafast raman loss
			spectroscopy (urls): instrumentation and principle},\ }\href
	{https://doi.org/https://doi.org/10.1002/jrs.2996} {\bibfield  {journal}
		{\bibinfo  {journal} {Journal of Raman Spectroscopy}\ }\textbf {\bibinfo
			{volume} {42}},\ \bibinfo {pages} {1883} (\bibinfo {year}
		{2011})}\BibitemShut {NoStop}%
	\bibitem [{\citenamefont {Dang}\ \emph {et~al.}(2011)\citenamefont {Dang},
		\citenamefont {Bolme}, \citenamefont {Moore},\ and\ \citenamefont
		{McGrane}}]{Dang_2011}%
	\BibitemOpen
	\bibfield  {author} {\bibinfo {author} {\bibfnamefont {N.~C.}\ \bibnamefont
			{Dang}}, \bibinfo {author} {\bibfnamefont {C.~A.}\ \bibnamefont {Bolme}},
		\bibinfo {author} {\bibfnamefont {D.~S.}\ \bibnamefont {Moore}},\ and\
		\bibinfo {author} {\bibfnamefont {S.~D.}\ \bibnamefont {McGrane}},\
	}\bibfield  {title} {\bibinfo {title} {Femtosecond stimulated raman
			scattering picosecond molecular thermometry in condensed phases},\ }\href
	{https://doi.org/10.1103/PhysRevLett.107.043001} {\bibfield  {journal}
		{\bibinfo  {journal} {Phys. Rev. Lett.}\ }\textbf {\bibinfo {volume} {107}},\
		\bibinfo {pages} {043001} (\bibinfo {year} {2011})}\BibitemShut {NoStop}%
	\bibitem [{\citenamefont {Harbola}\ \emph {et~al.}(2013)\citenamefont
		{Harbola}, \citenamefont {Umapathy},\ and\ \citenamefont
		{Mukamel}}]{Mukamel_2013}%
	\BibitemOpen
	\bibfield  {author} {\bibinfo {author} {\bibfnamefont {U.}~\bibnamefont
			{Harbola}}, \bibinfo {author} {\bibfnamefont {S.}~\bibnamefont {Umapathy}},\
		and\ \bibinfo {author} {\bibfnamefont {S.}~\bibnamefont {Mukamel}},\
	}\bibfield  {title} {\bibinfo {title} {Loss and gain signals in broadband
			stimulated-raman spectra: Theoretical analysis},\ }\href
	{https://doi.org/10.1103/PhysRevA.88.011801} {\bibfield  {journal} {\bibinfo
			{journal} {Phys. Rev. A}\ }\textbf {\bibinfo {volume} {88}},\ \bibinfo
		{pages} {011801(R)} (\bibinfo {year} {2013})}\BibitemShut {NoStop}%
	\bibitem [{\citenamefont {Lee}\ \emph {et~al.}(2004)\citenamefont {Lee},
		\citenamefont {Zhang}, \citenamefont {McCamant}, \citenamefont {Kukura},\
		and\ \citenamefont {Mathies}}]{Lee2004}%
	\BibitemOpen
	\bibfield  {author} {\bibinfo {author} {\bibfnamefont {S.}~\bibnamefont
			{Lee}}, \bibinfo {author} {\bibfnamefont {D.}~\bibnamefont {Zhang}}, \bibinfo
		{author} {\bibfnamefont {D.~W.}\ \bibnamefont {McCamant}}, \bibinfo {author}
		{\bibfnamefont {P.}~\bibnamefont {Kukura}},\ and\ \bibinfo {author}
		{\bibfnamefont {R.~A.}\ \bibnamefont {Mathies}},\ }\bibfield  {title}
	{\bibinfo {title} {Theory of femtosecond stimulated raman spectroscopy},\
	}\href {https://doi.org/10.1063/1.1777214} {\bibfield  {journal} {\bibinfo
			{journal} {J. Chem. Phys.}\ }\textbf {\bibinfo {volume} {121}},\ \bibinfo
		{pages} {3632} (\bibinfo {year} {2004})}\BibitemShut {NoStop}%
	\bibitem [{\citenamefont {Dorfman}\ \emph {et~al.}(2013)\citenamefont
		{Dorfman}, \citenamefont {Fingerhut},\ and\ \citenamefont
		{Mukamel}}]{Dorfman2013}%
	\BibitemOpen
	\bibfield  {author} {\bibinfo {author} {\bibfnamefont {K.~E.}\ \bibnamefont
			{Dorfman}}, \bibinfo {author} {\bibfnamefont {B.~P.}\ \bibnamefont
			{Fingerhut}},\ and\ \bibinfo {author} {\bibfnamefont {S.}~\bibnamefont
			{Mukamel}},\ }\bibfield  {title} {\bibinfo {title} {Time-resolved broadband
			raman spectroscopies: A unified six-wave-mixing representation},\ }\href
	{https://doi.org/10.1063/1.4821228} {\bibfield  {journal} {\bibinfo
			{journal} {J. Chem. Phys}\ }\textbf {\bibinfo {volume} {139}},\ \bibinfo
		{pages} {124113} (\bibinfo {year} {2013})}\BibitemShut {NoStop}%
	\bibitem [{\citenamefont {Potma}\ and\ \citenamefont
		{Mukamel}(2012)}]{Mukamel_Potma_CRS}%
	\BibitemOpen
	\bibfield  {author} {\bibinfo {author} {\bibfnamefont {E.~O.}\ \bibnamefont
			{Potma}}\ and\ \bibinfo {author} {\bibfnamefont {S.}~\bibnamefont
			{Mukamel}},\ }\bibfield  {title} {\bibinfo {title} {Theory of coherent raman
			scattering},\ }in\ \href@noop {} {\emph {\bibinfo {booktitle} {Coherent Raman
				Scattering Microscopy}}},\ \bibinfo {editor} {edited by\ \bibinfo {editor}
		{\bibfnamefont {J.-X.}\ \bibnamefont {Cheng}}\ and\ \bibinfo {editor}
		{\bibfnamefont {X.~S.}\ \bibnamefont {Xie}}}\ (\bibinfo  {publisher} {CRC
		Press},\ \bibinfo {address} {Boca Raton, FL},\ \bibinfo {year}
	{2012})\BibitemShut {NoStop}%
	\bibitem [{\citenamefont {Prince}\ \emph {et~al.}(2016)\citenamefont {Prince},
		\citenamefont {Frontiera},\ and\ \citenamefont {Potma}}]{Prince2016}%
	\BibitemOpen
	\bibfield  {author} {\bibinfo {author} {\bibfnamefont {R.~C.}\ \bibnamefont
			{Prince}}, \bibinfo {author} {\bibfnamefont {R.~R.}\ \bibnamefont
			{Frontiera}},\ and\ \bibinfo {author} {\bibfnamefont {E.~O.}\ \bibnamefont
			{Potma}},\ }\bibfield  {title} {\bibinfo {title} {Stimulated raman
			scattering: From bulk to nano},\ }\href
	{https://doi.org/10.1021/acs.chemrev.6b00545} {\bibfield  {journal} {\bibinfo
			{journal} {Chem. Rev.}\ }\textbf {\bibinfo {volume} {117}},\ \bibinfo
		{pages} {5070} (\bibinfo {year} {2016})}\BibitemShut {NoStop}%
	\bibitem [{\citenamefont {Kukura}\ \emph {et~al.}(2007)\citenamefont {Kukura},
		\citenamefont {McCamant},\ and\ \citenamefont {Mathies}}]{Mathies_review}%
	\BibitemOpen
	\bibfield  {author} {\bibinfo {author} {\bibfnamefont {P.}~\bibnamefont
			{Kukura}}, \bibinfo {author} {\bibfnamefont {D.~W.}\ \bibnamefont
			{McCamant}},\ and\ \bibinfo {author} {\bibfnamefont {R.~A.}\ \bibnamefont
			{Mathies}},\ }\bibfield  {title} {\bibinfo {title} {Femtosecond stimulated
			raman spectroscopy},\ }\href
	{https://doi.org/10.1146/annurev.physchem.58.032806.104456} {\bibfield
		{journal} {\bibinfo  {journal} {Annu. Rev. Phys. Chem.}\ }\textbf {\bibinfo
			{volume} {58}},\ \bibinfo {pages} {461} (\bibinfo {year} {2007})}\BibitemShut
	{NoStop}%
	\bibitem [{\citenamefont {Ferrante}\ \emph {et~al.}(2018)\citenamefont
		{Ferrante}, \citenamefont {Batignani}, \citenamefont {Fumero}, \citenamefont
		{Pontecorvo}, \citenamefont {Virga}, \citenamefont {Montemiglio},
		\citenamefont {Cerullo}, \citenamefont {Vos},\ and\ \citenamefont
		{Scopigno}}]{Ferrante2018}%
	\BibitemOpen
	\bibfield  {author} {\bibinfo {author} {\bibfnamefont {C.}~\bibnamefont
			{Ferrante}}, \bibinfo {author} {\bibfnamefont {G.}~\bibnamefont {Batignani}},
		\bibinfo {author} {\bibfnamefont {G.}~\bibnamefont {Fumero}}, \bibinfo
		{author} {\bibfnamefont {E.}~\bibnamefont {Pontecorvo}}, \bibinfo {author}
		{\bibfnamefont {A.}~\bibnamefont {Virga}}, \bibinfo {author} {\bibfnamefont
			{L.~C.}\ \bibnamefont {Montemiglio}}, \bibinfo {author} {\bibfnamefont
			{G.}~\bibnamefont {Cerullo}}, \bibinfo {author} {\bibfnamefont {M.~H.}\
			\bibnamefont {Vos}},\ and\ \bibinfo {author} {\bibfnamefont {T.}~\bibnamefont
			{Scopigno}},\ }\bibfield  {title} {\bibinfo {title} {Resonant broadband
			stimulated raman scattering in myoglobin},\ }\href
	{https://doi.org/10.1002/jrs.5323} {\bibfield  {journal} {\bibinfo  {journal}
			{J. Raman Spectrosc.}\ }\textbf {\bibinfo {volume} {49}},\ \bibinfo {pages}
		{913} (\bibinfo {year} {2018})}\BibitemShut {NoStop}%
	\bibitem [{\citenamefont {Ruhman}\ \emph {et~al.}(1988)\citenamefont {Ruhman},
		\citenamefont {Joly},\ and\ \citenamefont {Nelson}}]{Ruhman1988}%
	\BibitemOpen
	\bibfield  {author} {\bibinfo {author} {\bibfnamefont {S.}~\bibnamefont
			{Ruhman}}, \bibinfo {author} {\bibfnamefont {A.}~\bibnamefont {Joly}},\ and\
		\bibinfo {author} {\bibfnamefont {K.}~\bibnamefont {Nelson}},\ }\bibfield
	{title} {\bibinfo {title} {Coherent molecular vibrational motion observed in
			the time domain through impulsive stimulated raman scattering},\ }\href
	{https://doi.org/10.1109/3.146} {\bibfield  {journal} {\bibinfo  {journal}
			{{IEEE} J. Quant. Electron.}\ }\textbf {\bibinfo {volume} {24}},\ \bibinfo
		{pages} {460} (\bibinfo {year} {1988})}\BibitemShut {NoStop}%
	\bibitem [{\citenamefont {Kuramochi}\ and\ \citenamefont
		{Tahara}(2021)}]{Kuramochi2021}%
	\BibitemOpen
	\bibfield  {author} {\bibinfo {author} {\bibfnamefont {H.}~\bibnamefont
			{Kuramochi}}\ and\ \bibinfo {author} {\bibfnamefont {T.}~\bibnamefont
			{Tahara}},\ }\bibfield  {title} {\bibinfo {title} {Tracking ultrafast
			structural dynamics by time-domain raman spectroscopy},\ }\href
	{https://doi.org/10.1021/jacs.1c02545} {\bibfield  {journal} {\bibinfo
			{journal} {J. Am. Chem. Soc.}\ }\textbf {\bibinfo {volume} {143}},\ \bibinfo
		{pages} {9699} (\bibinfo {year} {2021})}\BibitemShut {NoStop}%
	\bibitem [{\citenamefont {Monacelli}\ \emph {et~al.}(2017)\citenamefont
		{Monacelli}, \citenamefont {Batignani}, \citenamefont {Fumero}, \citenamefont
		{Ferrante}, \citenamefont {Mukamel},\ and\ \citenamefont
		{Scopigno}}]{Monacelli2017}%
	\BibitemOpen
	\bibfield  {author} {\bibinfo {author} {\bibfnamefont {L.}~\bibnamefont
			{Monacelli}}, \bibinfo {author} {\bibfnamefont {G.}~\bibnamefont
			{Batignani}}, \bibinfo {author} {\bibfnamefont {G.}~\bibnamefont {Fumero}},
		\bibinfo {author} {\bibfnamefont {C.}~\bibnamefont {Ferrante}}, \bibinfo
		{author} {\bibfnamefont {S.}~\bibnamefont {Mukamel}},\ and\ \bibinfo {author}
		{\bibfnamefont {T.}~\bibnamefont {Scopigno}},\ }\bibfield  {title} {\bibinfo
		{title} {Manipulating impulsive stimulated raman spectroscopy with a chirped
			probe pulse},\ }\href {https://doi.org/10.1021/acs.jpclett.6b03027}
	{\bibfield  {journal} {\bibinfo  {journal} {J. Phys. Chem. Lett.}\ }\textbf
		{\bibinfo {volume} {8}},\ \bibinfo {pages} {966} (\bibinfo {year}
		{2017})}\BibitemShut {NoStop}%
	\bibitem [{\citenamefont {Batignani}\ \emph
		{et~al.}(2022{\natexlab{a}})\citenamefont {Batignani}, \citenamefont {Mai},
		\citenamefont {Fumero}, \citenamefont {Mukamel},\ and\ \citenamefont
		{Scopigno}}]{Batignani2022}%
	\BibitemOpen
	\bibfield  {author} {\bibinfo {author} {\bibfnamefont {G.}~\bibnamefont
			{Batignani}}, \bibinfo {author} {\bibfnamefont {E.}~\bibnamefont {Mai}},
		\bibinfo {author} {\bibfnamefont {G.}~\bibnamefont {Fumero}}, \bibinfo
		{author} {\bibfnamefont {S.}~\bibnamefont {Mukamel}},\ and\ \bibinfo {author}
		{\bibfnamefont {T.}~\bibnamefont {Scopigno}},\ }\bibfield  {title} {\bibinfo
		{title} {Absolute excited state molecular geometries revealed by resonance
			raman signals},\ }\bibfield  {journal} {\bibinfo  {journal} {Nat. Commun.}\
	}\textbf {\bibinfo {volume} {13}},\ \href
	{https://doi.org/10.1038/s41467-022-35099-3} {10.1038/s41467-022-35099-3}
	(\bibinfo {year} {2022}{\natexlab{a}})\BibitemShut {NoStop}%
	\bibitem [{\citenamefont {Liebel}\ \emph {et~al.}(2015)\citenamefont {Liebel},
		\citenamefont {Schnedermann}, \citenamefont {Wende},\ and\ \citenamefont
		{Kukura}}]{cit::IVS::kukura}%
	\BibitemOpen
	\bibfield  {author} {\bibinfo {author} {\bibfnamefont {M.}~\bibnamefont
			{Liebel}}, \bibinfo {author} {\bibfnamefont {C.}~\bibnamefont
			{Schnedermann}}, \bibinfo {author} {\bibfnamefont {T.}~\bibnamefont
			{Wende}},\ and\ \bibinfo {author} {\bibfnamefont {P.}~\bibnamefont
			{Kukura}},\ }\bibfield  {title} {\bibinfo {title} {Principles and
			applications of broadband impulsive vibrational spectroscopy},\ }\href
	{https://doi.org/10.1021/acs.jpca.5b05948} {\bibfield  {journal} {\bibinfo
			{journal} {J. Phys. Chem. A}\ }\textbf {\bibinfo {volume} {119}},\ \bibinfo
		{pages} {9506} (\bibinfo {year} {2015})}\BibitemShut {NoStop}%
	\bibitem [{\citenamefont {Gdor}\ \emph {et~al.}(2017)\citenamefont {Gdor},
		\citenamefont {Ghosh}, \citenamefont {Lioubashevski},\ and\ \citenamefont
		{Ruhman}}]{Gdor2017}%
	\BibitemOpen
	\bibfield  {author} {\bibinfo {author} {\bibfnamefont {I.}~\bibnamefont
			{Gdor}}, \bibinfo {author} {\bibfnamefont {T.}~\bibnamefont {Ghosh}},
		\bibinfo {author} {\bibfnamefont {O.}~\bibnamefont {Lioubashevski}},\ and\
		\bibinfo {author} {\bibfnamefont {S.}~\bibnamefont {Ruhman}},\ }\bibfield
	{title} {\bibinfo {title} {Nonresonant raman effects on femtosecond
			pump{\textendash}probe with chirped white light: Challenges and
			opportunities},\ }\href {https://doi.org/10.1021/acs.jpclett.7b00559}
	{\bibfield  {journal} {\bibinfo  {journal} {J. Phys. Chem. Lett.}\ }\textbf
		{\bibinfo {volume} {8}},\ \bibinfo {pages} {1920} (\bibinfo {year}
		{2017})}\BibitemShut {NoStop}%
	\bibitem [{\citenamefont {Batignani}\ \emph {et~al.}(2021)\citenamefont
		{Batignani}, \citenamefont {Sansone}, \citenamefont {Ferrante}, \citenamefont
		{Fumero}, \citenamefont {Mukamel},\ and\ \citenamefont
		{Scopigno}}]{Batignani_2021}%
	\BibitemOpen
	\bibfield  {author} {\bibinfo {author} {\bibfnamefont {G.}~\bibnamefont
			{Batignani}}, \bibinfo {author} {\bibfnamefont {C.}~\bibnamefont {Sansone}},
		\bibinfo {author} {\bibfnamefont {C.}~\bibnamefont {Ferrante}}, \bibinfo
		{author} {\bibfnamefont {G.}~\bibnamefont {Fumero}}, \bibinfo {author}
		{\bibfnamefont {S.}~\bibnamefont {Mukamel}},\ and\ \bibinfo {author}
		{\bibfnamefont {T.}~\bibnamefont {Scopigno}},\ }\bibfield  {title} {\bibinfo
		{title} {Excited-state energy surfaces in molecules revealed by impulsive
			stimulated raman excitation profiles},\ }\href
	{https://doi.org/10.1021/acs.jpclett.1c02209} {\bibfield  {journal} {\bibinfo
			{journal} {J. Phys. Chem. Lett.}\ }\textbf {\bibinfo {volume} {12}},\
		\bibinfo {pages} {9239} (\bibinfo {year} {2021})}\BibitemShut {NoStop}%
	\bibitem [{\citenamefont {Yoshizawa}\ \emph {et~al.}(1994)\citenamefont
		{Yoshizawa}, \citenamefont {Hattori},\ and\ \citenamefont
		{Kobayashi}}]{Yoshizawa1994}%
	\BibitemOpen
	\bibfield  {author} {\bibinfo {author} {\bibfnamefont {M.}~\bibnamefont
			{Yoshizawa}}, \bibinfo {author} {\bibfnamefont {Y.}~\bibnamefont {Hattori}},\
		and\ \bibinfo {author} {\bibfnamefont {T.}~\bibnamefont {Kobayashi}},\
	}\bibfield  {title} {\bibinfo {title} {Femtosecond time-resolved resonance
			raman gain spectroscopy in polydiacetylene},\ }\href
	{https://doi.org/10.1103/PhysRevB.49.13259} {\bibfield  {journal} {\bibinfo
			{journal} {Phys. Rev. B}\ }\textbf {\bibinfo {volume} {49}},\ \bibinfo
		{pages} {13259} (\bibinfo {year} {1994})}\BibitemShut {NoStop}%
	\bibitem [{\citenamefont {Yoshizawa}\ and\ \citenamefont
		{Kurosawa}(1999)}]{Yoshizawa1999}%
	\BibitemOpen
	\bibfield  {author} {\bibinfo {author} {\bibfnamefont {M.}~\bibnamefont
			{Yoshizawa}}\ and\ \bibinfo {author} {\bibfnamefont {M.}~\bibnamefont
			{Kurosawa}},\ }\bibfield  {title} {\bibinfo {title} {Femtosecond
			time-resolved raman spectroscopy using stimulated raman scattering},\
	}\bibfield  {journal} {\bibinfo  {journal} {Phys. Rev. A}\ }\textbf {\bibinfo
		{volume} {61}},\ \href {https://doi.org/10.1103/PhysRevA.61.013808}
	{10.1103/PhysRevA.61.013808} (\bibinfo {year} {1999})\BibitemShut {NoStop}%
	\bibitem [{\citenamefont {Takeuchi}\ \emph {et~al.}(2008)\citenamefont
		{Takeuchi}, \citenamefont {Ruhman}, \citenamefont {Tsuneda}, \citenamefont
		{Chiba}, \citenamefont {Taketsugu},\ and\ \citenamefont
		{Tahara}}]{Takeuchi2008}%
	\BibitemOpen
	\bibfield  {author} {\bibinfo {author} {\bibfnamefont {S.}~\bibnamefont
			{Takeuchi}}, \bibinfo {author} {\bibfnamefont {S.}~\bibnamefont {Ruhman}},
		\bibinfo {author} {\bibfnamefont {T.}~\bibnamefont {Tsuneda}}, \bibinfo
		{author} {\bibfnamefont {M.}~\bibnamefont {Chiba}}, \bibinfo {author}
		{\bibfnamefont {T.}~\bibnamefont {Taketsugu}},\ and\ \bibinfo {author}
		{\bibfnamefont {T.}~\bibnamefont {Tahara}},\ }\bibfield  {title} {\bibinfo
		{title} {Spectroscopic tracking of structural evolution in ultrafast stilbene
			photoisomerization},\ }\href {https://doi.org/10.1126/science.1160902}
	{\bibfield  {journal} {\bibinfo  {journal} {Science}\ }\textbf {\bibinfo
			{volume} {322}},\ \bibinfo {pages} {1073} (\bibinfo {year}
		{2008})}\BibitemShut {NoStop}%
	\bibitem [{\citenamefont {Dietze}\ and\ \citenamefont
		{Mathies}(2016)}]{Dietze2016}%
	\BibitemOpen
	\bibfield  {author} {\bibinfo {author} {\bibfnamefont {D.~R.}\ \bibnamefont
			{Dietze}}\ and\ \bibinfo {author} {\bibfnamefont {R.~A.}\ \bibnamefont
			{Mathies}},\ }\bibfield  {title} {\bibinfo {title} {Femtosecond stimulated
			raman spectroscopy},\ }\href {https://doi.org/10.1002/cphc.201600104}
	{\bibfield  {journal} {\bibinfo  {journal} {{ChemPhysChem}}\ }\textbf
		{\bibinfo {volume} {17}},\ \bibinfo {pages} {1224} (\bibinfo {year}
		{2016})}\BibitemShut {NoStop}%
	\bibitem [{\citenamefont {Mukamel}\ and\ \citenamefont
		{Biggs}(2011)}]{Biggs2011}%
	\BibitemOpen
	\bibfield  {author} {\bibinfo {author} {\bibfnamefont {S.}~\bibnamefont
			{Mukamel}}\ and\ \bibinfo {author} {\bibfnamefont {J.~D.}\ \bibnamefont
			{Biggs}},\ }\bibfield  {title} {\bibinfo {title} {Communication: Comment on
			the effective temporal and spectral resolution of impulsive stimulated raman
			signals},\ }\href {https://doi.org/10.1063/1.3581889} {\bibfield  {journal}
		{\bibinfo  {journal} {J. Chem. Phys.}\ }\textbf {\bibinfo {volume} {134}},\
		\bibinfo {pages} {161101} (\bibinfo {year} {2011})}\BibitemShut {NoStop}%
	\bibitem [{\citenamefont {Provencher}\ \emph {et~al.}(2014)\citenamefont
		{Provencher}, \citenamefont {B{\'{e}}rub{\'{e}}}, \citenamefont {Parker},
		\citenamefont {Greetham}, \citenamefont {Towrie}, \citenamefont {Hellmann},
		\citenamefont {C{\^{o}}t{\'{e}}}, \citenamefont {Stingelin}, \citenamefont
		{Silva},\ and\ \citenamefont {Hayes}}]{Provencher2014}%
	\BibitemOpen
	\bibfield  {author} {\bibinfo {author} {\bibfnamefont {F.}~\bibnamefont
			{Provencher}}, \bibinfo {author} {\bibfnamefont {N.}~\bibnamefont
			{B{\'{e}}rub{\'{e}}}}, \bibinfo {author} {\bibfnamefont {A.~W.}\ \bibnamefont
			{Parker}}, \bibinfo {author} {\bibfnamefont {G.~M.}\ \bibnamefont
			{Greetham}}, \bibinfo {author} {\bibfnamefont {M.}~\bibnamefont {Towrie}},
		\bibinfo {author} {\bibfnamefont {C.}~\bibnamefont {Hellmann}}, \bibinfo
		{author} {\bibfnamefont {M.}~\bibnamefont {C{\^{o}}t{\'{e}}}}, \bibinfo
		{author} {\bibfnamefont {N.}~\bibnamefont {Stingelin}}, \bibinfo {author}
		{\bibfnamefont {C.}~\bibnamefont {Silva}},\ and\ \bibinfo {author}
		{\bibfnamefont {S.~C.}\ \bibnamefont {Hayes}},\ }\bibfield  {title} {\bibinfo
		{title} {Direct observation of ultrafast long-range charge separation at
			polymer{\textendash}fullerene heterojunctions},\ }\bibfield  {journal}
	{\bibinfo  {journal} {Nat. Commun.}\ }\textbf {\bibinfo {volume} {5}},\ \href
	{https://doi.org/10.1038/ncomms5288} {10.1038/ncomms5288} (\bibinfo {year}
	{2014})\BibitemShut {NoStop}%
	\bibitem [{\citenamefont {Batignani}\ \emph {et~al.}(2015)\citenamefont
		{Batignani}, \citenamefont {Bossini}, \citenamefont {Palo}, \citenamefont
		{Ferrante}, \citenamefont {Pontecorvo}, \citenamefont {Cerullo},
		\citenamefont {Kimel},\ and\ \citenamefont {Scopigno}}]{Batignani2015}%
	\BibitemOpen
	\bibfield  {author} {\bibinfo {author} {\bibfnamefont {G.}~\bibnamefont
			{Batignani}}, \bibinfo {author} {\bibfnamefont {D.}~\bibnamefont {Bossini}},
		\bibinfo {author} {\bibfnamefont {N.~D.}\ \bibnamefont {Palo}}, \bibinfo
		{author} {\bibfnamefont {C.}~\bibnamefont {Ferrante}}, \bibinfo {author}
		{\bibfnamefont {E.}~\bibnamefont {Pontecorvo}}, \bibinfo {author}
		{\bibfnamefont {G.}~\bibnamefont {Cerullo}}, \bibinfo {author} {\bibfnamefont
			{A.}~\bibnamefont {Kimel}},\ and\ \bibinfo {author} {\bibfnamefont
			{T.}~\bibnamefont {Scopigno}},\ }\bibfield  {title} {\bibinfo {title}
		{Probing ultrafast photo-induced dynamics of the exchange energy in a
			heisenberg antiferromagnet},\ }\href
	{https://doi.org/10.1038/nphoton.2015.121} {\bibfield  {journal} {\bibinfo
			{journal} {Nat. Photonics}\ }\textbf {\bibinfo {volume} {9}},\ \bibinfo
		{pages} {506} (\bibinfo {year} {2015})}\BibitemShut {NoStop}%
	\bibitem [{\citenamefont {Zhou}\ \emph {et~al.}(2015)\citenamefont {Zhou},
		\citenamefont {Yu},\ and\ \citenamefont {Bragg}}]{Zhou2015}%
	\BibitemOpen
	\bibfield  {author} {\bibinfo {author} {\bibfnamefont {J.}~\bibnamefont
			{Zhou}}, \bibinfo {author} {\bibfnamefont {W.}~\bibnamefont {Yu}},\ and\
		\bibinfo {author} {\bibfnamefont {A.~E.}\ \bibnamefont {Bragg}},\ }\bibfield
	{title} {\bibinfo {title} {Structural relaxation of photoexcited
			quaterthiophenes probed with vibrational specificity},\ }\href
	{https://doi.org/10.1021/acs.jpclett.5b01472} {\bibfield  {journal} {\bibinfo
			{journal} {J. Phys. Chem. Lett.}\ }\textbf {\bibinfo {volume} {6}},\
		\bibinfo {pages} {3496} (\bibinfo {year} {2015})}\BibitemShut {NoStop}%
	\bibitem [{\citenamefont {Batignani}\ \emph
		{et~al.}(2016{\natexlab{a}})\citenamefont {Batignani}, \citenamefont
		{Pontecorvo}, \citenamefont {Ferrante}, \citenamefont {Aschi}, \citenamefont
		{Elles},\ and\ \citenamefont {Scopigno}}]{Batignani2016}%
	\BibitemOpen
	\bibfield  {author} {\bibinfo {author} {\bibfnamefont {G.}~\bibnamefont
			{Batignani}}, \bibinfo {author} {\bibfnamefont {E.}~\bibnamefont
			{Pontecorvo}}, \bibinfo {author} {\bibfnamefont {C.}~\bibnamefont
			{Ferrante}}, \bibinfo {author} {\bibfnamefont {M.}~\bibnamefont {Aschi}},
		\bibinfo {author} {\bibfnamefont {C.~G.}\ \bibnamefont {Elles}},\ and\
		\bibinfo {author} {\bibfnamefont {T.}~\bibnamefont {Scopigno}},\ }\bibfield
	{title} {\bibinfo {title} {Visualizing excited-state dynamics of a diaryl
			thiophene: Femtosecond stimulated raman scattering as a probe of conjugated
			molecules},\ }\href {https://doi.org/10.1021/acs.jpclett.6b01137} {\bibfield
		{journal} {\bibinfo  {journal} {J. Phys. Chem. Lett.}\ }\textbf {\bibinfo
			{volume} {7}},\ \bibinfo {pages} {2981} (\bibinfo {year}
		{2016}{\natexlab{a}})}\BibitemShut {NoStop}%
	\bibitem [{\citenamefont {Barclay}\ \emph {et~al.}(2017)\citenamefont
		{Barclay}, \citenamefont {Quincy}, \citenamefont {Williams-Young},
		\citenamefont {Caricato},\ and\ \citenamefont {Elles}}]{Barclay2017}%
	\BibitemOpen
	\bibfield  {author} {\bibinfo {author} {\bibfnamefont {M.~S.}\ \bibnamefont
			{Barclay}}, \bibinfo {author} {\bibfnamefont {T.~J.}\ \bibnamefont {Quincy}},
		\bibinfo {author} {\bibfnamefont {D.~B.}\ \bibnamefont {Williams-Young}},
		\bibinfo {author} {\bibfnamefont {M.}~\bibnamefont {Caricato}},\ and\
		\bibinfo {author} {\bibfnamefont {C.~G.}\ \bibnamefont {Elles}},\ }\bibfield
	{title} {\bibinfo {title} {Accurate assignments of excited-state resonance
			raman spectra: A benchmark study combining experiment and theory},\ }\href
	{https://doi.org/10.1021/acs.jpca.7b09467} {\bibfield  {journal} {\bibinfo
			{journal} {J. Phys. Chem. AJ. Phys. Chem. A}\ }\textbf {\bibinfo {volume}
			{121}},\ \bibinfo {pages} {7937} (\bibinfo {year} {2017})}\BibitemShut
	{NoStop}%
	\bibitem [{\citenamefont {Quincy}\ \emph {et~al.}(2018)\citenamefont {Quincy},
		\citenamefont {Barclay}, \citenamefont {Caricato},\ and\ \citenamefont
		{Elles}}]{Quincy2018}%
	\BibitemOpen
	\bibfield  {author} {\bibinfo {author} {\bibfnamefont {T.~J.}\ \bibnamefont
			{Quincy}}, \bibinfo {author} {\bibfnamefont {M.~S.}\ \bibnamefont {Barclay}},
		\bibinfo {author} {\bibfnamefont {M.}~\bibnamefont {Caricato}},\ and\
		\bibinfo {author} {\bibfnamefont {C.~G.}\ \bibnamefont {Elles}},\ }\bibfield
	{title} {\bibinfo {title} {Probing dynamics in higher-lying electronic states
			with resonance-enhanced femtosecond stimulated raman spectroscopy},\ }\href
	{https://doi.org/10.1021/acs.jpca.8b07855} {\bibfield  {journal} {\bibinfo
			{journal} {J. Phys. Chem. A}\ }\textbf {\bibinfo {volume} {122}},\ \bibinfo
		{pages} {8308} (\bibinfo {year} {2018})}\BibitemShut {NoStop}%
	\bibitem [{\citenamefont {Takaya}\ \emph {et~al.}(2018)\citenamefont {Takaya},
		\citenamefont {Anan},\ and\ \citenamefont {Iwata}}]{Takaya2018}%
	\BibitemOpen
	\bibfield  {author} {\bibinfo {author} {\bibfnamefont {T.}~\bibnamefont
			{Takaya}}, \bibinfo {author} {\bibfnamefont {M.}~\bibnamefont {Anan}},\ and\
		\bibinfo {author} {\bibfnamefont {K.}~\bibnamefont {Iwata}},\ }\bibfield
	{title} {\bibinfo {title} {Vibrational relaxation dynamics of
			$\beta$-carotene and its derivatives with substituents on terminal rings in
			electronically excited states as studied by femtosecond time-resolved
			stimulated raman spectroscopy in the near-{IR} region},\ }\href
	{https://doi.org/10.1039/C7CP06343A} {\bibfield  {journal} {\bibinfo
			{journal} {Phys. Chem. Chem. Phys.}\ }\textbf {\bibinfo {volume} {20}},\
		\bibinfo {pages} {3320} (\bibinfo {year} {2018})}\BibitemShut {NoStop}%
	\bibitem [{\citenamefont {Hontani}\ \emph {et~al.}(2018)\citenamefont
		{Hontani}, \citenamefont {Kloz}, \citenamefont {Pol{\'{\i}}vka},
		\citenamefont {Shukla}, \citenamefont {Sobotka},\ and\ \citenamefont
		{Kennis}}]{Hontani2018}%
	\BibitemOpen
	\bibfield  {author} {\bibinfo {author} {\bibfnamefont {Y.}~\bibnamefont
			{Hontani}}, \bibinfo {author} {\bibfnamefont {M.}~\bibnamefont {Kloz}},
		\bibinfo {author} {\bibfnamefont {T.}~\bibnamefont {Pol{\'{\i}}vka}},
		\bibinfo {author} {\bibfnamefont {M.~K.}\ \bibnamefont {Shukla}}, \bibinfo
		{author} {\bibfnamefont {R.}~\bibnamefont {Sobotka}},\ and\ \bibinfo {author}
		{\bibfnamefont {J.~T.~M.}\ \bibnamefont {Kennis}},\ }\bibfield  {title}
	{\bibinfo {title} {Molecular origin of photoprotection in cyanobacteria
			probed by watermarked femtosecond stimulated raman spectroscopy},\ }\href
	{https://doi.org/10.1021/acs.jpclett.8b00663} {\bibfield  {journal} {\bibinfo
			{journal} {J. Phys. Chem. Lett.}\ }\textbf {\bibinfo {volume} {9}},\
		\bibinfo {pages} {1788} (\bibinfo {year} {2018})}\BibitemShut {NoStop}%
	\bibitem [{\citenamefont {Piontkowski}\ and\ \citenamefont
		{McCamant}(2018)}]{Piontkowski2018}%
	\BibitemOpen
	\bibfield  {author} {\bibinfo {author} {\bibfnamefont {Z.}~\bibnamefont
			{Piontkowski}}\ and\ \bibinfo {author} {\bibfnamefont {D.~W.}\ \bibnamefont
			{McCamant}},\ }\bibfield  {title} {\bibinfo {title} {Excited-state
			planarization in donor-bridge dye sensitizers: Phenylene versus thiophene
			bridges},\ }\href {https://doi.org/10.1021/jacs.8b05463} {\bibfield
		{journal} {\bibinfo  {journal} {J. Am. Chem. Soc.}\ }\textbf {\bibinfo
			{volume} {140}},\ \bibinfo {pages} {11046} (\bibinfo {year}
		{2018})}\BibitemShut {NoStop}%
	\bibitem [{\citenamefont {Fang}\ \emph {et~al.}(2019)\citenamefont {Fang},
		\citenamefont {Tang},\ and\ \citenamefont {Chen}}]{Fang2019}%
	\BibitemOpen
	\bibfield  {author} {\bibinfo {author} {\bibfnamefont {C.}~\bibnamefont
			{Fang}}, \bibinfo {author} {\bibfnamefont {L.}~\bibnamefont {Tang}},\ and\
		\bibinfo {author} {\bibfnamefont {C.}~\bibnamefont {Chen}},\ }\bibfield
	{title} {\bibinfo {title} {{Unveiling coupled electronic and vibrational
				motions of chromophores in condensed phases}},\ }\href
	{https://doi.org/10.1063/1.5128388} {\bibfield  {journal} {\bibinfo
			{journal} {The Journal of Chemical Physics}\ }\textbf {\bibinfo {volume}
			{151}},\ \bibinfo {pages} {200901} (\bibinfo {year} {2019})}\BibitemShut
	{NoStop}%
	\bibitem [{\citenamefont {Fujisawa}\ \emph {et~al.}(2016)\citenamefont
		{Fujisawa}, \citenamefont {Kuramochi}, \citenamefont {Hosoi}, \citenamefont
		{Takeuchi},\ and\ \citenamefont {Tahara}}]{Fujisawa2016}%
	\BibitemOpen
	\bibfield  {author} {\bibinfo {author} {\bibfnamefont {T.}~\bibnamefont
			{Fujisawa}}, \bibinfo {author} {\bibfnamefont {H.}~\bibnamefont {Kuramochi}},
		\bibinfo {author} {\bibfnamefont {H.}~\bibnamefont {Hosoi}}, \bibinfo
		{author} {\bibfnamefont {S.}~\bibnamefont {Takeuchi}},\ and\ \bibinfo
		{author} {\bibfnamefont {T.}~\bibnamefont {Tahara}},\ }\bibfield  {title}
	{\bibinfo {title} {Role of coherent low-frequency motion in excited-state
			proton transfer of green fluorescent protein studied by time-resolved
			impulsive stimulated raman spectroscopy},\ }\href
	{https://doi.org/10.1021/jacs.5b11038} {\bibfield  {journal} {\bibinfo
			{journal} {J. Am. Chem. Soc.}\ }\textbf {\bibinfo {volume} {138}},\ \bibinfo
		{pages} {3942} (\bibinfo {year} {2016})}\BibitemShut {NoStop}%
	\bibitem [{\citenamefont {Kim}\ \emph {et~al.}(2020)\citenamefont {Kim},
		\citenamefont {Kim}, \citenamefont {Kang}, \citenamefont {Hong},
		\citenamefont {Würthner},\ and\ \citenamefont {Kim}}]{Kim2020}%
	\BibitemOpen
	\bibfield  {author} {\bibinfo {author} {\bibfnamefont {W.}~\bibnamefont
			{Kim}}, \bibinfo {author} {\bibfnamefont {T.}~\bibnamefont {Kim}}, \bibinfo
		{author} {\bibfnamefont {S.}~\bibnamefont {Kang}}, \bibinfo {author}
		{\bibfnamefont {Y.}~\bibnamefont {Hong}}, \bibinfo {author} {\bibfnamefont
			{F.}~\bibnamefont {Würthner}},\ and\ \bibinfo {author} {\bibfnamefont
			{D.}~\bibnamefont {Kim}},\ }\bibfield  {title} {\bibinfo {title} {Tracking
			structural evolution during symmetry-breaking charge separation in
			quadrupolar perylene bisimide with time-resolved impulsive stimulated raman
			spectroscopy},\ }\href
	{https://doi.org/https://doi.org/10.1002/anie.202002733} {\bibfield
		{journal} {\bibinfo  {journal} {Angew. Chem.}\ }\textbf {\bibinfo {volume}
			{59}},\ \bibinfo {pages} {8571} (\bibinfo {year} {2020})}\BibitemShut
	{NoStop}%
	\bibitem [{\citenamefont {Kim}\ and\ \citenamefont {Musser}(2021)}]{Kim2021}%
	\BibitemOpen
	\bibfield  {author} {\bibinfo {author} {\bibfnamefont {W.}~\bibnamefont
			{Kim}}\ and\ \bibinfo {author} {\bibfnamefont {A.~J.}\ \bibnamefont
			{Musser}},\ }\bibfield  {title} {\bibinfo {title} {Tracking ultrafast
			reactions in organic materials through vibrational coherence: vibronic
			coupling mechanisms in singlet fission},\ }\bibfield  {journal} {\bibinfo
		{journal} {Adv. Phys.: X}\ }\textbf {\bibinfo {volume} {6}},\ \href
	{https://doi.org/10.1080/23746149.2021.1918022}
	{10.1080/23746149.2021.1918022} (\bibinfo {year} {2021})\BibitemShut
	{NoStop}%
	\bibitem [{\citenamefont {Borrego-Varillas}\ \emph {et~al.}(2021)\citenamefont
		{Borrego-Varillas}, \citenamefont {Nenov}, \citenamefont {Kabaci{\'{n}}ski},
		\citenamefont {Conti}, \citenamefont {Ganzer}, \citenamefont {Oriana},
		\citenamefont {Jaiswal}, \citenamefont {Delfino}, \citenamefont {Weingart},
		\citenamefont {Manzoni}, \citenamefont {Rivalta}, \citenamefont {Garavelli},\
		and\ \citenamefont {Cerullo}}]{BorregoVarillas2021}%
	\BibitemOpen
	\bibfield  {author} {\bibinfo {author} {\bibfnamefont {R.}~\bibnamefont
			{Borrego-Varillas}}, \bibinfo {author} {\bibfnamefont {A.}~\bibnamefont
			{Nenov}}, \bibinfo {author} {\bibfnamefont {P.}~\bibnamefont
			{Kabaci{\'{n}}ski}}, \bibinfo {author} {\bibfnamefont {I.}~\bibnamefont
			{Conti}}, \bibinfo {author} {\bibfnamefont {L.}~\bibnamefont {Ganzer}},
		\bibinfo {author} {\bibfnamefont {A.}~\bibnamefont {Oriana}}, \bibinfo
		{author} {\bibfnamefont {V.~K.}\ \bibnamefont {Jaiswal}}, \bibinfo {author}
		{\bibfnamefont {I.}~\bibnamefont {Delfino}}, \bibinfo {author} {\bibfnamefont
			{O.}~\bibnamefont {Weingart}}, \bibinfo {author} {\bibfnamefont
			{C.}~\bibnamefont {Manzoni}}, \bibinfo {author} {\bibfnamefont
			{I.}~\bibnamefont {Rivalta}}, \bibinfo {author} {\bibfnamefont
			{M.}~\bibnamefont {Garavelli}},\ and\ \bibinfo {author} {\bibfnamefont
			{G.}~\bibnamefont {Cerullo}},\ }\bibfield  {title} {\bibinfo {title}
		{Tracking excited state decay mechanisms of pyrimidine nucleosides in real
			time},\ }\bibfield  {journal} {\bibinfo  {journal} {Nat. Commun.}\ }\textbf
	{\bibinfo {volume} {12}},\ \href {https://doi.org/10.1038/s41467-021-27535-7}
	{10.1038/s41467-021-27535-7} (\bibinfo {year} {2021})\BibitemShut {NoStop}%
	\bibitem [{\citenamefont {Tanimura}\ and\ \citenamefont
		{Mukamel}(1993)}]{Tanimura1993}%
	\BibitemOpen
	\bibfield  {author} {\bibinfo {author} {\bibfnamefont {Y.}~\bibnamefont
			{Tanimura}}\ and\ \bibinfo {author} {\bibfnamefont {S.}~\bibnamefont
			{Mukamel}},\ }\bibfield  {title} {\bibinfo {title} {Two-dimensional
			femtosecond vibrational spectroscopy of liquids},\ }\href
	{https://doi.org/10.1063/1.465484} {\bibfield  {journal} {\bibinfo  {journal}
			{J. Chem. Phys}\ }\textbf {\bibinfo {volume} {99}},\ \bibinfo {pages} {9496}
		(\bibinfo {year} {1993})}\BibitemShut {NoStop}%
	\bibitem [{\citenamefont {Batignani}\ \emph
		{et~al.}(2016{\natexlab{b}})\citenamefont {Batignani}, \citenamefont
		{Pontecorvo}, \citenamefont {Giovannetti}, \citenamefont {Ferrante},
		\citenamefont {Fumero},\ and\ \citenamefont
		{Scopigno}}]{Batignani2016SciRep}%
	\BibitemOpen
	\bibfield  {author} {\bibinfo {author} {\bibfnamefont {G.}~\bibnamefont
			{Batignani}}, \bibinfo {author} {\bibfnamefont {E.}~\bibnamefont
			{Pontecorvo}}, \bibinfo {author} {\bibfnamefont {G.}~\bibnamefont
			{Giovannetti}}, \bibinfo {author} {\bibfnamefont {C.}~\bibnamefont
			{Ferrante}}, \bibinfo {author} {\bibfnamefont {G.}~\bibnamefont {Fumero}},\
		and\ \bibinfo {author} {\bibfnamefont {T.}~\bibnamefont {Scopigno}},\
	}\bibfield  {title} {\bibinfo {title} {Electronic resonances in broadband
			stimulated raman spectroscopy},\ }\bibfield  {journal} {\bibinfo  {journal}
		{Sci. Rep.}\ }\textbf {\bibinfo {volume} {6}},\ \href
	{https://doi.org/10.1038/srep18445} {10.1038/srep18445} (\bibinfo {year}
	{2016}{\natexlab{b}})\BibitemShut {NoStop}%
	\bibitem [{\citenamefont {Batignani}\ \emph {et~al.}(2020)\citenamefont
		{Batignani}, \citenamefont {Ferrante},\ and\ \citenamefont
		{Scopigno}}]{Batignani2020}%
	\BibitemOpen
	\bibfield  {author} {\bibinfo {author} {\bibfnamefont {G.}~\bibnamefont
			{Batignani}}, \bibinfo {author} {\bibfnamefont {C.}~\bibnamefont
			{Ferrante}},\ and\ \bibinfo {author} {\bibfnamefont {T.}~\bibnamefont
			{Scopigno}},\ }\bibfield  {title} {\bibinfo {title} {Accessing excited state
			molecular vibrations by femtosecond stimulated raman spectroscopy},\ }\href
	{https://doi.org/10.1021/acs.jpclett.0c01971} {\bibfield  {journal} {\bibinfo
			{journal} {J. Phys. Chem. Lett.}\ }\textbf {\bibinfo {volume} {11}},\
		\bibinfo {pages} {7805} (\bibinfo {year} {2020})}\BibitemShut {NoStop}%
	\bibitem [{\citenamefont {Mukamel}(1995)}]{cit::Mukamel}%
	\BibitemOpen
	\bibfield  {author} {\bibinfo {author} {\bibfnamefont {S.}~\bibnamefont
			{Mukamel}},\ }\href@noop {} {\emph {\bibinfo {title} {Principles of Nonlinear
				Spectroscopy}}}\ (\bibinfo  {publisher} {Oxford University Press},\ \bibinfo
	{address} {New York},\ \bibinfo {year} {1995})\BibitemShut {NoStop}%
	\bibitem [{\citenamefont {Batignani}\ \emph
		{et~al.}(2022{\natexlab{b}})\citenamefont {Batignani}, \citenamefont
		{Fumero}, \citenamefont {Mai}, \citenamefont {Martinati},\ and\ \citenamefont
		{Scopigno}}]{batignani_OMX}%
	\BibitemOpen
	\bibfield  {author} {\bibinfo {author} {\bibfnamefont {G.}~\bibnamefont
			{Batignani}}, \bibinfo {author} {\bibfnamefont {G.}~\bibnamefont {Fumero}},
		\bibinfo {author} {\bibfnamefont {E.}~\bibnamefont {Mai}}, \bibinfo {author}
		{\bibfnamefont {M.}~\bibnamefont {Martinati}},\ and\ \bibinfo {author}
		{\bibfnamefont {T.}~\bibnamefont {Scopigno}},\ }\bibfield  {title} {\bibinfo
		{title} {Stimulated raman lineshapes in the large light–matter interaction
			limit},\ }\href {https://doi.org/https://doi.org/10.1016/j.omx.2021.100134}
	{\bibfield  {journal} {\bibinfo  {journal} {Opt. Mater. X}\ }\textbf
		{\bibinfo {volume} {13}},\ \bibinfo {pages} {100134} (\bibinfo {year}
		{2022}{\natexlab{b}})}\BibitemShut {NoStop}%
	\bibitem [{\citenamefont {Loudon}(2000)}]{cit::LoudonBook}%
	\BibitemOpen
	\bibfield  {author} {\bibinfo {author} {\bibnamefont {Loudon}},\ }\href@noop
	{} {\emph {\bibinfo {title} {The Quantum Theory of Light}}}\ (\bibinfo
	{publisher} {Oxford University Press},\ \bibinfo {year} {2000})\BibitemShut
	{NoStop}%
	\bibitem [{\citenamefont {Polli}\ \emph {et~al.}(2018)\citenamefont {Polli},
		\citenamefont {Kumar}, \citenamefont {Valensise}, \citenamefont {Marangoni},\
		and\ \citenamefont {Cerullo}}]{Polli2018}%
	\BibitemOpen
	\bibfield  {author} {\bibinfo {author} {\bibfnamefont {D.}~\bibnamefont
			{Polli}}, \bibinfo {author} {\bibfnamefont {V.}~\bibnamefont {Kumar}},
		\bibinfo {author} {\bibfnamefont {C.~M.}\ \bibnamefont {Valensise}}, \bibinfo
		{author} {\bibfnamefont {M.}~\bibnamefont {Marangoni}},\ and\ \bibinfo
		{author} {\bibfnamefont {G.}~\bibnamefont {Cerullo}},\ }\bibfield  {title}
	{\bibinfo {title} {Broadband coherent raman scattering microscopy},\ }\href
	{https://doi.org/10.1002/lpor.201800020} {\bibfield  {journal} {\bibinfo
			{journal} {Laser Photonics Rev.}\ }\textbf {\bibinfo {volume} {12}},\
		\bibinfo {pages} {1800020} (\bibinfo {year} {2018})}\BibitemShut {NoStop}%
	\bibitem [{\citenamefont {Virga}\ \emph {et~al.}(2019)\citenamefont {Virga},
		\citenamefont {Ferrante}, \citenamefont {Batignani}, \citenamefont {Fazio},
		\citenamefont {Nunn}, \citenamefont {Ferrari}, \citenamefont {Cerullo},\ and\
		\citenamefont {Scopigno}}]{Virga2019}%
	\BibitemOpen
	\bibfield  {author} {\bibinfo {author} {\bibfnamefont {A.}~\bibnamefont
			{Virga}}, \bibinfo {author} {\bibfnamefont {C.}~\bibnamefont {Ferrante}},
		\bibinfo {author} {\bibfnamefont {G.}~\bibnamefont {Batignani}}, \bibinfo
		{author} {\bibfnamefont {D.~D.}\ \bibnamefont {Fazio}}, \bibinfo {author}
		{\bibfnamefont {A.~D.~G.}\ \bibnamefont {Nunn}}, \bibinfo {author}
		{\bibfnamefont {A.~C.}\ \bibnamefont {Ferrari}}, \bibinfo {author}
		{\bibfnamefont {G.}~\bibnamefont {Cerullo}},\ and\ \bibinfo {author}
		{\bibfnamefont {T.}~\bibnamefont {Scopigno}},\ }\bibfield  {title} {\bibinfo
		{title} {Coherent anti-stokes raman spectroscopy of single and multi-layer
			graphene},\ }\bibfield  {journal} {\bibinfo  {journal} {Nat. Commun.}\
	}\textbf {\bibinfo {volume} {10}},\ \href
	{https://doi.org/10.1038/s41467-019-11165-1} {10.1038/s41467-019-11165-1}
	(\bibinfo {year} {2019})\BibitemShut {NoStop}%
	\bibitem [{\citenamefont {Dahl}\ and\ \citenamefont
		{Springborg}(1988)}]{Dahl1988}%
	\BibitemOpen
	\bibfield  {author} {\bibinfo {author} {\bibfnamefont {J.~P.}\ \bibnamefont
			{Dahl}}\ and\ \bibinfo {author} {\bibfnamefont {M.}~\bibnamefont
			{Springborg}},\ }\bibfield  {title} {\bibinfo {title} {{The Morse oscillator
				in position space, momentum space, and phase space}},\ }\href
	{https://doi.org/10.1063/1.453761} {\bibfield  {journal} {\bibinfo  {journal}
			{The Journal of Chemical Physics}\ }\textbf {\bibinfo {volume} {88}},\
		\bibinfo {pages} {4535} (\bibinfo {year} {1988})}\BibitemShut {NoStop}%
	\bibitem [{\citenamefont {Ferrante}\ \emph {et~al.}(2016)\citenamefont
		{Ferrante}, \citenamefont {Pontecorvo}, \citenamefont {Cerullo},
		\citenamefont {Vos},\ and\ \citenamefont {Scopigno}}]{Ferrante2016}%
	\BibitemOpen
	\bibfield  {author} {\bibinfo {author} {\bibfnamefont {C.}~\bibnamefont
			{Ferrante}}, \bibinfo {author} {\bibfnamefont {E.}~\bibnamefont
			{Pontecorvo}}, \bibinfo {author} {\bibfnamefont {G.}~\bibnamefont {Cerullo}},
		\bibinfo {author} {\bibfnamefont {M.~H.}\ \bibnamefont {Vos}},\ and\ \bibinfo
		{author} {\bibfnamefont {T.}~\bibnamefont {Scopigno}},\ }\bibfield  {title}
	{\bibinfo {title} {Direct observation of subpicosecond vibrational dynamics
			in photoexcited myoglobin},\ }\href {https://doi.org/10.1038/nchem.2569}
	{\bibfield  {journal} {\bibinfo  {journal} {Nat. Chem.}\ }\textbf {\bibinfo
			{volume} {8}},\ \bibinfo {pages} {1137} (\bibinfo {year} {2016})}\BibitemShut
	{NoStop}%
	\bibitem [{\citenamefont {Ferrante}\ \emph {et~al.}(2020)\citenamefont
		{Ferrante}, \citenamefont {Batignani}, \citenamefont {Pontecorvo},
		\citenamefont {Montemiglio}, \citenamefont {Vos},\ and\ \citenamefont
		{Scopigno}}]{Ferrante2020}%
	\BibitemOpen
	\bibfield  {author} {\bibinfo {author} {\bibfnamefont {C.}~\bibnamefont
			{Ferrante}}, \bibinfo {author} {\bibfnamefont {G.}~\bibnamefont {Batignani}},
		\bibinfo {author} {\bibfnamefont {E.}~\bibnamefont {Pontecorvo}}, \bibinfo
		{author} {\bibfnamefont {L.~C.}\ \bibnamefont {Montemiglio}}, \bibinfo
		{author} {\bibfnamefont {M.~H.}\ \bibnamefont {Vos}},\ and\ \bibinfo {author}
		{\bibfnamefont {T.}~\bibnamefont {Scopigno}},\ }\bibfield  {title} {\bibinfo
		{title} {Ultrafast dynamics and vibrational relaxation in six-coordinate heme
			proteins revealed by femtosecond stimulated raman spectroscopy},\ }\href
	{https://doi.org/10.1021/jacs.9b10560} {\bibfield  {journal} {\bibinfo
			{journal} {J. Am. Chem. Soc.}\ }\textbf {\bibinfo {volume} {142}},\ \bibinfo
		{pages} {2285} (\bibinfo {year} {2020})}\BibitemShut {NoStop}%
\end{thebibliography}
%

\end{document}